\documentclass[fleqn,usenatbib]{mnras}

\usepackage{newtxtext,newtxmath}

\usepackage[T1]{fontenc}

\DeclareRobustCommand{\VAN}[3]{#2}
\let\VANthebibliography\thebibliography
\def\thebibliography{\DeclareRobustCommand{\VAN}[3]{##3}\VANthebibliography}

\usepackage{graphicx}	
\usepackage{amsmath}	
\usepackage{xcolor}     
\usepackage{enumerate}
\usepackage{mathtools}

\newcommand{\molh}{\mathrm{H}_2}

\title[Semi-analytic modeling of Pop III halos]{A self-consistent semi-analytic model for Population III star formation in minihalos}

\author[Hegde \& Furlanetto]{
Sahil Hegde$^{1}$\thanks{E-mail: sahil@astro.ucla.edu (SH)} and
Steven R. Furlanetto$^{1}$
\\
$^{1}$ Department of Physics \& Astronomy, University of California, Los Angeles, 475 Portola Plaza, Los Angeles, CA 90095, USA\\
}

\date{Accepted XXX. Received YYY; in original form ZZZ}

\pubyear{2023}

\begin{document}
\label{firstpage}
\pagerange{\pageref{firstpage}--\pageref{lastpage}}
\maketitle

\begin{abstract}
The formation of the first stars marks a watershed moment in the history of our universe. As the first luminous structures, these stars (also known as Population III, or Pop III stars) seed the first galaxies and begin the process of reionization. We construct an analytic model to self-consistently trace the formation of Pop III stars inside minihalos in the presence of the fluctuating ultraviolet background, relic dark matter-baryon relative velocities from the early universe, and an X-ray background, which largely work to suppress cooling of gas and delay the formation of this first generation of stars. We demonstrate the utility of this framework in a semi-analytic model for early star formation that also follows the transition between Pop III and Pop II star formation inside these halos. Using our new prescription for the criteria allowing Pop III star formation, we follow a population of dark matter halos from $z=50$ through $z=6$ and examine the global star formation history, finding that each process defines its own key epoch: \emph{(i)} the stream velocity dominates at the highest redshifts ($z\gtrsim30$), \emph{(ii)} the UV background sets the tone at intermediate times ($30\gtrsim z\gtrsim15$), and \emph{(iii)} X-rays control the end of Pop III star formation at the latest times ($z\lesssim 15$). In all of our models, Pop III stars continue to form down to $z\sim 7-10$, when their supernovae will be potentially observable with forthcoming instruments. Finally, we identify the signatures of variations in the Pop III physics in the global 21-cm spin-flip signal of atomic hydrogen.
\end{abstract}


\section{Introduction}
Formed in the pristine hydrogen and helium gas of the early universe, the first stars, also known as Population III (Pop III) stars, are thought to have been very different from the stars we observe today (Pop I/II) (\citealp{Bromm13}, \citealp{LF13}). In these metal-free clouds, Pop III stars must rely on the radiative transitions of molecular hydrogen ($\molh$) to  cool, fragment, and collapse to high densities. However, this cooling is comparatively inefficient (relative to atomic hydrogen cooling at high temperatures or cooling in metal-enriched gas via transitions of, e.g., CO, H$_2$O, etc.) and is believed to result in a population of very massive stars. These stars are expected to form in so-called dark matter (DM) `minihalos' of $10^5-10^6\ M_\odot$ and likely formed in isolation or groups of a few (\citealp{Abel02}, \citealp{Bromm02}, \citealp{Bromm13}). At such large masses, these stars will have short ($\sim 5$ Myr) lives, some of which may have ended their lives in superluminous pair-instability supernovae (SNe; \citealp{Barkat67}, \citealp{Fryer01}, \citealp{HegerWoosley02}, \citealp{Heger03}). In so doing, they ejected metals into the interstellar and intergalactic media, paving the way for future generations of metal-enriched star formation (\citealp{Ferrara00}, \citealp{Madau01}, \citealp{Furlanetto03}).

Observationally probing the epoch of the first stars, which is colloquially referred to as `Cosmic Dawn', is a difficult endeavor. Given their short lifetimes and their formation in small groups, Pop III stars and their host minihalos were likely very faint \citep{Mebane18}. In addition, these halos likely only dominated in the very early universe, requiring observations that can probe to $z\gtrsim 20-30$. To that end, observations of the indirect signatures of these stars, such as in their SNe or through the highly redshifted 21-cm line, offer a far more promising avenue to study Cosmic Dawn (e.g., \citealp{Mebane20}, \citealp{Magg22}, \citealp{Munoz22}, \citealp{LazarBromm22}). However, even the most promising of these avenues, observations of the 21-cm line, will be challenging. Therefore, leveraging robust theory models will be crucial to the design of instruments that are sensitive to the relevant areas of parameter space and development of software pipelines to robustly infer those parameters. 

Pop III halos have been studied theoretically with a variety of methods (in order of decreasing computational cost): numerical simulations (e.g., \citealp{Machacek01}, \citealp{Abel02}, \citealp{WiseAbel07}, \citealp{OsheaNorman08}, \citealp{Maio10}, \citealp{Stacy12}, \citealp{Hirano15}, \citealp{Xu16}, \citealp{Sarmento18}, \citealp{Park21}, \citealp{Kulkarni21}, \citealp{Schauer21}), semi-analytic models (e.g., \citealp{Trenti09}, \citealp{Jaacks18}, \citealp{Visbal18}, \citealp{Mebane18}, \citealp{Visbal20}, \citealp{Magg22}), and analytic calculations (e.g., \citealp{Tegmark1997}, \citealp{Haiman00}, \citealp{McKeeTan08}, \citealp{Kulkarni13}, \citealp{Ricotti16}). 
Because the models are essentially unconstrained in the absence of observations, exploring the breadth of the available parameter space is challenging, especially with detailed numerical calculations. Semi-analytic models offer a compromise between the computational efficiency of an analytic calculation and the rigor of a numerical simulation. In practice, applications of these models to high redshift star formation, such as the model described in this work, begin with a numerical model for dark matter (DM) halo growth (such as through abundance matching to a simulated halo mass function or through simulated halo merger trees) and fold in the physics of star formation (initial mass function, supernovae, feedback, etc.) through analytic approximations. In such models, these two pieces are connected with the minimum DM halo mass for Pop III stars to form. 

The minimum mass is the host halo mass scale above which a star forming cloud is able to form, cool, and collapse, driven by the presence of $\molh$ (\citealp{Haiman96b}, \citealp{Haiman96a}, \citealp{Tegmark1997}, \citealp{Machacek01}). As a result, an understanding of processes that affect the amount of $\molh$ present in a halo is crucial for understanding when and where early star formation can occur. There are several key processes believed to significantly affect the $\molh$ fraction in a halo---the relic velocity between baryons and DM in the early universe (environmental) and the buildup of X-ray and photodissociating UV backgrounds (feedback) (e.g., \citealp{LF13}, \citealp{Ricotti16}, \citealp{Mebane18}, \citealp{Schauer21}, \citealp{Kulkarni21}, \citealp{Munoz22}, \citealp{Nebrin23}). 
 
The effect of a strong UV background, specifically in the Lyman-Werner (LW) bands of $\molh$ (11.2-13.6 eV), is relatively well-studied, both analytically and in numerical simulations (e.g., \citealp{Tegmark1997}, \citealp{Machacek01}, \citealp{SBH10}, \citealp{WG11}, \citealp{Visbal14}, \citealp{Kulkarni21}). As stars form, they radiate and build up a metagalactic LW background, which breaks down $\molh$ and suppresses later generations of star formation. The relative DM-baryon motion, often referred to as the \textit{stream velocity}, has only come to be thought of as significant in recent years and, as such, is relatively less studied (e.g., \citealp{TH10}, \citealp{Dalal10},\citealp{Maio11_streaming}, \citealp{Naoz13}, \citealp{Fialkov14}). The stream velocity suppresses accretion and limits the gas fraction in the least massive halos. As such, these two processes both negatively impact the $\molh$ content of a halo. X-rays, on the other hand, are more complicated, as they can both positively and negatively feed back into the star formation process through photoionization and heating of the IGM, respectively (e.g., \citealp{Machacek03}, \citealp{Ricotti04}, \citealp{Hummel15}, \citealp{Ricotti16}, \citealp{Park21}). Each of these mechanisms can be parameterized in terms of the minimum DM halo mass for star formation, and most work has focused on studying these processes independently. The recent simulations of \citet{Kulkarni21} and \citet{Schauer21} are some of the first attempts to study the joint effects of the two negative processes, the LW photodissociation and stream velocity, but present discrepant estimates for this minimum mass scale.  
 
Motivated by these uncertainties and the lack of a single model that encompasses all of these effects,\footnote{We note, however, that while this paper was in the final stages of preparation, \citet{Nebrin23} presented a calculation for the minimum halo mass to host Pop III star formation that included the LW background and streaming mechanisms, in addition to several others with more modest effects, but not the effects of X-rays. We will compare to their model throughout.} in this work we present an analytic calculation for the minimum star-forming halo mass that incorporates the aforementioned three processes. We update the semi-analytic model presented in \citet{Mebane18} to explore the effects of variations in the underlying Pop III physics on the global star formation rate density. Finally, we use these results to make preliminary observational predictions, specifically in calculating Pop III SN rates and the global 21-cm signal.

The paper is structured as follows. In \S~\ref{sec:min_mass} we outline our analytic calculation of the minimum mass and present a fitting formula that summarizes these calculations. In \S~\ref{sec:semianalytic_model}, we describe the semi-analytic model and our improvements to \citet{Mebane18}. In \S~\ref{sec:results} we present our results---namely the fiducial model and the effects of varying the physics incorporated into the minimum mass. In \S~\ref{sec:comparison} we compare our results to previous works and in \S~\ref{sec:observations} we present potential observable signatures of Pop III star formation. Finally, in \S~\ref{sec:conclusions}, we summarize our main conclusions.  

In this work we use a flat $\Lambda$CDM cosmology with $\Omega_{\rm m} = 0.3111$, $\Omega_{\Lambda} = 0.6889$, $\Omega_{\rm b} = 0.0489$, $\sigma_8 = 0.8102$, $n_s = 0.9665$, and $h = 0.6766$, consistent with the results of \citet{Planck21}.

\section{The Minimum Mass for Pop III Star Formation}\label{sec:min_mass}
In this section, we will describe the components of our simple analytic model of the aforementioned processes---the metagalactic LW background, stream velocity, and X-ray background---to generate an  estimate of the minimum star-forming halo mass scale. Schematically, this is broken down as follows: a halo must first accrete gas (filter mass; \S~\ref{sssec:filter_mass}) and then must be able to efficiently cool (cooling threshold; \S~\ref{sssec:cooling_threshold}). These two thresholds are modified in the presence of a DM-baryon relative velocity (Sections~\ref{sssec:filter_w_stream} and~\ref{sssec:thermalization_threshold}). Once the first stars form, they produce a UV background (dissociation threshold; \S~\ref{sssec:LW_bckd}) and an X-ray background (\S~\ref{sssec:xray-bkgd}) that can affect the minimum mass for subsequent generations of stars.

\subsection{Accretion mass}\label{sssec:filter_mass}

In the absence of both a LW background and a stream velocity, there are two relevant processes --- a halo must be able to accrete baryons and must then be able to cool and form stars. The former of these is governed by the so-called `filtering' scale, which is the scale below which baryonic perturbations are suppressed relative to the DM fluctuations \citep{Gnedin98} by thermal pressure. This scale, or the associated mass scale (the \textit{filter mass}), is sometimes referred to as the \textit{cosmological Jeans mass} because it resembles a time-averaged Jeans mass and effectively sets a minimum halo mass necessary for gas accretion.

\subsubsection{No stream velocity}
Building from the calculation of the filtering scale outlined in \citet{Naoz07}, \citet{Naoz13} calculate the filter mass using numerical simulations and show that the results can be reproduced with an analytic calculation from linear theory. Inspired by the calculations of \citet{Naoz13}, in this section we describe a simple model that produces remarkable agreement with the analytic model therein. 

Following the discussion in \citet{Gnedin00}, we calculate the filtering wavenumber, $k_F$, defined as the scale at which the baryonic perturbations grow significantly compared to the DM fluctuations. Assuming that the perturbations satisfy $\delta_b(t,k=0) = \delta_c(t, k=0)$, the ratio of these quantities can be expanded as
\begin{equation}
    \frac{\delta_b}{\delta_c} \approx 1-\frac{k^2}{k_F^2}
\end{equation}
where we have dropped the higher order terms and assumed that the perturbations are small.
If we introduce the growth factor $D(t)$ and define $A(t)\equiv D(t)/k_F^2$, this becomes
\begin{equation}\label{eq:baryon_rat_D}
    \frac{\delta_b}{\delta_c} = 1-\frac{A(t)}{D(t)}k^2
\end{equation}
The linearized Euler equation can be recast in terms of the density perturbations as
\begin{equation}\label{eq:euler}
    \frac{\partial^2\delta}{\partial t^2} + 2H\frac{\partial \delta}{\partial t} = 4\pi G\bar{\rho}\delta - \frac{c_s^2k^2}{a^2}\delta
\end{equation}
Substituting eq.~\ref{eq:euler} for baryon and DM overdensities into eq.~\ref{eq:baryon_rat_D} yields
\begin{equation}\label{eq:A_diffeq}
    \frac{d^2A}{dt^2} + 2H\frac{dA}{dt} = \frac{c_s^2}{a^2}D(t)
\end{equation}
which, when solved, gives the evolution of $A(t)$, or, equivalently, the filtering scale $k_F(a)$. Noting that the Jeans wavenumber can be written as $k_J^2 = (3/2) \Omega_m H^2 a^2/c_s^2$, the full solution of eq.~\ref{eq:A_diffeq} can be cast in terms of $k_F^2$:
\begin{equation}\label{eq:full_soln}
    \frac{1}{k_F^2} = \frac{1}{D(t)}\int_0^tdt' a^2(t')\frac{\ddot{D}(t')+2H(t')\dot{D}(t')}{k_J^2(t')}\int_{t'}^{t''}\frac{dt''}{a^2(t'')}
\end{equation}
For large redshifts ($z \gtrsim 5$; the relevant regime here), we know that $\Omega_m\to 1$, so we can use the results for an Einstein-de Sitter cosmology; i.e., that $a\propto t^{2/3}$ and $D\propto a$. With these, eq.~\ref{eq:full_soln} simplifies to 
\begin{equation}\label{eq:kF_soln}
    \frac{1}{k_F^2(t)} = \frac{3}{a}\int_0^a\frac{da'}{k_J^2(a')}\bigg(1-\sqrt{\frac{a'}{a}}\bigg)
\end{equation}

\subsubsection{Including the DM-baryon relative velocity}\label{sssec:filter_w_stream}
To this point, we have ignored the effects of a relative velocity between DM and baryons. However, \citet{TH10} showed that the relative velocity between DM and baryon density fluctuations following recombination can have important consequences for structure formation in the early universe. This relative velocity, often referred to as the \textit{stream velocity}, is a product of the differing growth histories of DM and baryon perturbations prior to recombination and decoupling. That is, because they only interact gravitationally, DM particles began collapsing into their potential wells well before the end of recombination. Baryons, on the other hand, were prevented from such collapse by the thermal pressure that resulted from their coupling to the photon field. Once recombination had concluded, the photons and baryons decoupled and the baryons were free to fall into DM potential wells. However, they retained memory of their motion prior to that point and a spatially-varying distribution of relative velocities was produced as a result. It has since been shown, analytically and with numerical simulations, that this relative \textit{stream velocity} has a variety of effects on structure formation in the early universe, ranging from reducing the number density of DM halos to suppressing the gas content of halos (e.g., \citealp{Dalal10}, \citealp{TH10}, \citealp{OlearyMcquinn12}, \citealp{McQuinnOleary12}, \citealp{Naoz12}, \citealp{Naoz13}, \citealp{Fialkov14}, \citealp{Williams22}, \citealp{Lake23}). It is the latter phenomenon that is especially relevant for Pop III star formation, as a suppression of gas accretion can delay $\molh$ cooling and collapse.\footnote{The suppression of halo number density has important implications for structure formation, especially at the highest redshifts, when we expect the very first stars to form. However, we defer analysis of this contribution to future work and here limit our focus to the effect of the stream velocity on the gas being used to form stars.}

\citet{Naoz13} carry out the full self-consistent calculation of this effect, which requires modification of the definition of the filtering scale to account for large scale structure and inclusion of the temperature and density fluctuations in the coupled differential equations describing the growth of perturbations (i.e., eq.~\ref{eq:euler}). As a simple approximation to this calculation, we propose taking the stream velocity as a modification to the sound speed (which appears only in calculation of the Jeans wavenumber) and adding it in quadrature \citep{Stacy11}:
\begin{equation}
    c_s\mapsto\sqrt{A c_s^2 + v_\mathrm{bc}^2}
\end{equation}
where we include the constant $A$ to account for differences between our calculation of the sound speed and that of \citet{Naoz13}. Calibrating our calculation to the results of \citet{Naoz13}, we set $A = 0.64$. 

\begin{figure}
    \centering
\includegraphics[width=\columnwidth]{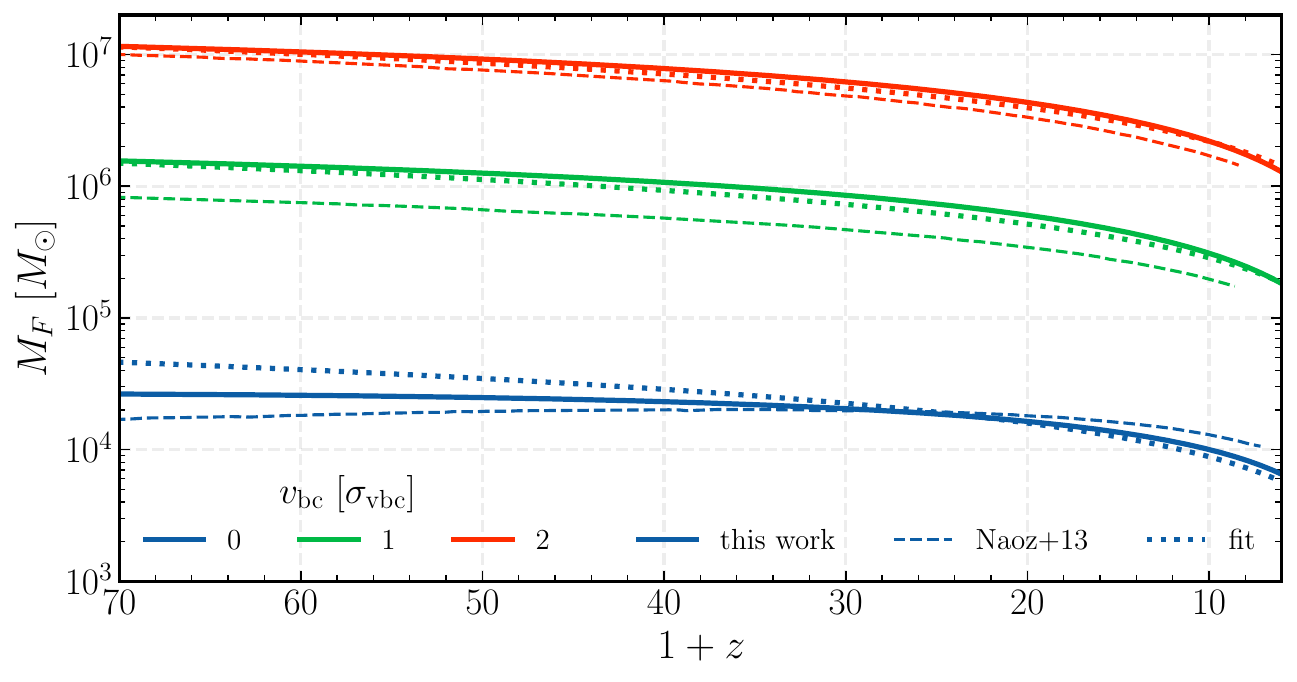}
    \caption{The filter mass calculated using our approximation (solid) compared to the results of \citet{Naoz13} (dashed) and the fit to our calculation (eq.~\ref{eq:filter_threshold}; dotted) for various values of the stream velocity (different colors).}
    \label{fig:Mf_full_compare}
\end{figure}

\begin{figure}
    \centering
    \includegraphics[width=\columnwidth]{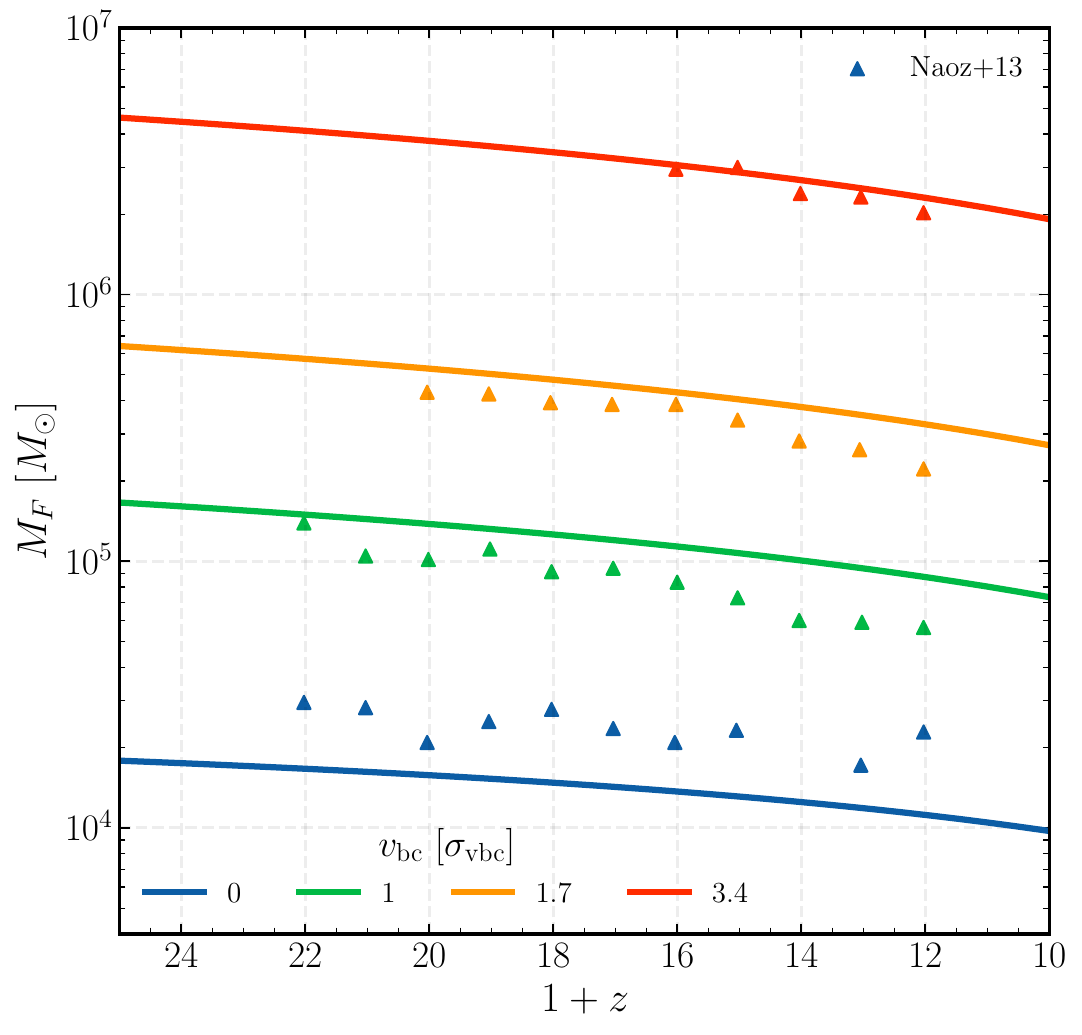}
    \caption{The filter mass calculated using our approximation (solid curves) compared to the numerically simulated results of \citet{Naoz13} (points) for various values of the stream velocity (different colors).}
    \label{fig:Mf_numeric_compare}
\end{figure}

In Figure~\ref{fig:Mf_full_compare}, we present the results of this calculation for various values of the stream velocity.\footnote{We conventionally refer to the stream velocity in multiples of the root-mean-square value of its Maxwellian distribution, which is $\sigma_{\rm vbc} = 30\ {\rm km\ s^{-1}}$ at $z_{\rm rec} = 1100$.} While our results do not precisely agree with the analytic calculation of \citet{Naoz13}, the approximation we employ performs remarkably well, as our estimates of the filter mass are discrepant by less than a factor of 2 and the agreement improves with increasing $v_{\rm bc}$. In Figure~\ref{fig:Mf_numeric_compare}, we compare our analytic approximation to the numerical simulations of \citet{Naoz13}, demonstrating similar agreement to that seen in Figure~\ref{fig:Mf_full_compare}. Note that to compare the results of this three-dimensional analytic calculation to the simulated results, we scale the velocity by a factor of $1/\sqrt{3}$ because the stream velocity acts in a particular direction, whereas the sound speed is isotropic.

We find that the following fitting formula reproduces our treatment of the filter mass quite well:
\begin{equation}\label{eq:filter_threshold}
    M_F(v_{\rm bc}, z) \simeq 1.66\times 10^4 \bigg(1+\frac{v_{\rm bc}}{\sigma_{\rm vbc}}\bigg)^{5.02}\bigg(\frac{1+z}{21}\bigg)^{0.85}\ M_\odot
\end{equation}

\subsection{Molecular hydrogen cooling threshold}\label{sssec:cooling_threshold}
 Once a halo is able to accrete gas, we must then find the threshold mass necessary to build up sufficient molecular hydrogen so that the gas can cool and collapse to high densities. Molecular hydrogen is primarily formed through a two step process, as follows:
\begin{align}\label{eq:H2_formation}
    {\rm H} + {\rm e}^- &\to {\rm H}^- + h\nu \ (k_9) \\
    {\rm H}^- + {\rm H} &\to {\rm H}_2 + {\rm e}^-\ (k_{10})
\end{align}
where the associated reaction rate coefficients $k_i$ are given in \citet{SBH10}. Note that for this calculation we use the T4 spectrum rate coefficients given therein. Accounting for the formation of $\molh$ via these two processes and the destruction of ${\rm H}^-$ by CMB photons (with associated rate coefficient $k_{25}$), the net production rate coefficient of $\molh$ is
\begin{equation}
    k_\mathrm{form} = k_9\bigg[\frac{k_{10}n_{\rm H}}{k_{10}n_{\rm H} + k_{25}}\bigg],
\end{equation}
where $n_H$ is the hydrogen number density and we have limited our focus to $\molh$ formation in the gas-phase.\footnote{Once a halo has been enriched by early generations of star formation, $\molh$ formation could, in principle, be catalyzed by the presence of dust grains \citep{Nakatani20}. However, for $T_{\rm vir} \leq 10^4\ {\rm K}$, where $\molh$ cooling will be dominant, \citet{Nebrin23} find that the dust-catalyzed formation rate will be smaller than the gas-phase formation rate except in the cases of a highly-enriched halo forming $\molh$ with maximal efficiency. \citet{Yamaguchi23} find that the volume-filling fraction of metal-enriched wind bubbles will be $\lesssim10$\% and the average metallicity of the universe $\langle Z\rangle/Z_\odot \lesssim 10^{-3}$ for $z\gtrsim 6$ in both their optimistic Pop III and Pop II models. Therefore, even if a significant number of halos are forming Pop II stars, it is unlikely that these halos will be able to enrich nearby Pop III halos to a degree that dust catalysis will be the dominant $\molh$ formation channel.}

The halo density is therefore a key input to the cooling process. Motivated by their numerical simulations, \citet{Visbal14b} argued that high-$z$ minihaloes fall into two regimes. At very small masses, the maximum gas density in a halo follows $n_H \propto T_{\rm vir}^{3/2}$, which reflects the maximum temperature allowed by adiabatic compression from the ambient IGM temperature to the halo virial temperature. For these small haloes, the IGM entropy is sufficiently large, compared to the entropy generated during halo collapse, that this adiabatic limit is a good approximation. However, at larger halo masses the entropy is dominated by halo formation, and in this regime the gas settles into a ``universal'' profile in which the central core density is independent of halo mass (at a fixed redshift). To normalize the scalings in these two regimes, we fit to the gas densities found by \citet{OlearyMcquinn12}, which yields\footnote{We have chosen to normalize to these results, which fall a factor of a few below that of \citet{Visbal14b}, in order to estimate the \textit{typical} halo gas density rather than the maximum. Variations in this normalization are effectively folded into the choice of our free parameter, $\zeta$ (introduced later). We discuss the effect of making a different density choice (e.g., \citealp{Nebrin23}) in more detail in \S~\ref{ssec:model_comparison}.}
\begin{equation}\label{eq:ngas}
    n_{\rm H} \simeq \begin{dcases}
        6.19\ \bigg(\frac{T_{\rm vir}}{10^3\ {\rm K}}\bigg)\ \mathrm{cm}^{-3}, &{\rm for}\ T_{\rm vir} < 2\times 10^3\ {\rm K} \\
      12.38\ \bigg(\frac{1+z}{21}\bigg)^3\ \mathrm{cm}^{-3}, &{\rm for}\ T_{\rm vir} \geq 2\times 10^3\ {\rm K} .       
    \end{dcases}
\end{equation}
We note that the transition between these regimes occurs at a redshift-dependent halo mass, \begin{equation}\label{eq:M_turn}
    M_{\rm turn} = 9.64\times 10^5 \bigg(\frac{1+z}{21}\bigg)^{-3/2}\ M_\odot.
\end{equation}

Defining the ionized fraction $x_{\rm HII} \equiv n_{\rm HII}/n_{\rm H}$ and molecular fraction $f_{\molh} \equiv n_{\molh}/n_{\rm H}$, the above pair of reactions (eq.~\ref{eq:H2_formation}) yield the evolution equations
\begin{align}\label{eq:H_rate_eqns}
    \dot{x}_{\rm HII} &= -\alpha_B n_{\rm H} x_{\rm HII}^2 \\ 
    \dot{f}_{\molh} &= k_{\rm form}n_{\rm H}x_{\rm HII}(1-x_{\rm HII} - 2f_{\molh})
\end{align}
where we have introduced the Case B recombination coefficient $\alpha_B$ to account for electron depletion due to recombination with H atoms. 

Following the discussion outlined in \citet{Tegmark1997}, in the limit of inefficient cooling (where $n$ and $T$ are constant), these can be solved to find that
\begin{equation}
    f_{\molh}(t) - f_{\molh}^i \approx \frac{k_{\rm form}}{\alpha_B}\ln(1+t/t_{\rm rec}^i)
\end{equation}
where $t_{\rm rec}^i \equiv (\alpha_B n_{\rm H} x_{\rm HII})^{-1}$ is the initial recombination timescale. For $t\ll t_{\rm rec}^i$, the density of electrons is sufficiently large that $\molh$ is produced at a constant rate $k_{\rm form}/\alpha_B$. For $t\gg t_{\rm rec}^i$, electrons will have been sufficiently depleted and the $\molh$ fraction will grow slowly. Therefore, when $t\sim t_{\rm rec}^i$, the $\molh$ fraction will reach a saturation level
\begin{equation}\label{eq:molH_frac}
    f_{\molh,\rm  sat} \approx \frac{k_{\rm form}}{\alpha_B} \approx 4.97\times 10^{-4} \bigg(\frac{T}{10^3\ {\rm K}}\bigg)^{1.52}
\end{equation}
which is a factor of $\sim 1.5$ larger than the value reported in \citet{Tegmark1997}, where the difference results from the incorrect use of the Case A recombination coefficient in that calculation (as noted by \citealp{Nebrin23}).\footnote{To calculate the second equality, we have approximated $k_{\rm form}\approx k_9$ (because $k_{25}\ll nk_{10}$ for the gas densities of relevance here) and made a power-law fit to the Case B recombination coefficient given in \citet{Draine11} between $30\ {\rm K} \leq T \leq 10^4\ {\rm K}$.}

With this in hand, we need to find the critical level of $\molh$ that needs to build up for cooling to become efficient. This can be found by comparing the cooling time of a cloud of gas to (a fraction $\zeta$ of) the then-current Hubble time; i.e., $t_{\rm cool} < \zeta t_H$. The cooling time is given by
\begin{equation}\label{eq:cool_time}
    t_{\rm cool} = \frac{3k_B T_{\rm vir}}{2\Lambda(n_{\rm H}, T_{\rm vir}) f_{\molh}}
\end{equation}
where for the molecular hydrogen cooling function we use the approximation to that found by \citet{GP98}, valid between the temperatures $120\ {\rm K}$ and $6400\ {\rm K}$ \citep{TS09}:
\begin{equation}
    \Lambda(n_{\rm H}, T) \simeq 10^{-31.6}\bigg(\frac{T}{100\ {\rm K}}\bigg)^{3.4}\bigg(\frac{n_{\rm H}}{10^{-4}{\rm cm^{-3}}}\bigg)\ {\rm erg\ s^{-1}}
\end{equation}
For the era of the first stars, when $z\gg 1$, we can again approximately use the results for an Einstein-de Sitter cosmology, for which the Hubble time is
\begin{equation}\label{eq:t_h}
    t_H = \frac{2}{3H_0}\Big[\Omega_{m}(1+z)^3\Big]^{-1/2} \approx 6.52\times 10^{9}\ \Big[\Omega_{m}h^2(1+z)^3\Big]^{-1/2}\ \mathrm{yr} 
\end{equation}
Enforcing the condition that cooling must occur on timescales shorter than a fraction $\zeta$ of the Hubble time (meant to represent the rate at which a halo accumulates thermal energy through accretion), we find that the critical $\molh$ density needed for efficient cooling at a virial temperature $T$ is
\begin{equation}\label{eq:nH2_crit}
    n_{\molh, \rm crit} = 1.534\times 10^{-4} \zeta^{-1}\big(\Omega_m h^2\big)^{1/2}\bigg(\frac{T}{10^3\ {\rm K}}\bigg)^{-2.4}\bigg(\frac{1+z}{21}\bigg)^{3/2}\ {\rm cm^{-3}}
\end{equation}
From this, we can compute the critical $\molh$ fraction as $f_{\molh, \rm crit} = n_{\molh, \rm crit}/n_{\rm H}$. With the virial temperature given in \citet{BL01} and plugging in the values for our chosen cosmology, this can be written as
\begin{equation}\label{eq:fH2_crit}
    f_{\molh, \rm crit} \simeq 8.17\times 10^{-7} \zeta^{-1}\bigg(\frac{M}{10^6 M_\odot}\bigg)^{-2.27}\bigg(\frac{1+z}{21}\bigg)^{-1.9}
\end{equation}
It turns out that the critical $\molh$ fraction will be achieved in the low-mass regime (in eq.~\ref{eq:ngas}) for the redshifts of interest, so we have omitted the solution associated with the high-mass halo density. We will see that once we introduce a photodissociating Lyman-Werner background, as is discussed in \S~\ref{sssec:LW_bckd}, the high-mass regime will become important. 

Comparing this threshold to the molecular hydrogen fraction in a halo (eq.~\ref{eq:molH_frac}), we can solve for the critical virial temperature (or mass) necessary for efficient cooling (setting $\zeta = 0.25$ for our fiducial calculations; see \citealp{Visbal14}). This yields the cooling threshold
\begin{equation}\label{eq:cooling_threshold}
    M_{\rm cool}(z) \simeq 1.55\times 10^{5}\bigg(\frac{\zeta}{0.25}\bigg)^{-0.3}\bigg(\frac{1+z}{21}\bigg)^{-1}\ M_\odot
\end{equation}
This threshold (along with the filter mass; see \S~\ref{sssec:filter_mass}) sets a baseline value for the minimum halo mass for efficient cooling in the absence of any radiation backgrounds or other external effects and is displayed in the flat part of the curves in Figure~\ref{fig:mcrit_jlw}. That is, for each $z$, there is a critical value of the Lyman-Werner background intensity below which the cooling threshold sets the minimum mass.

\subsection{The photodissociating Lyman-Werner background}\label{sssec:LW_bckd}
Once the first stars form, they produce radiation backgrounds that make subsequent generations of star formation more complex by affecting the amount of $\molh$ available to cool. The primary process that suppresses the $\molh$ content of a halo is negative feedback associated with the Lyman-Werner background; i.e. radiation of UV photons with energies of 11.2-13.6 eV that can dissociate molecular hydrogen. Therefore, as more stars form, a growing LW background builds up (e.g., \citealp{Visbal14}) and future generations of star formation are delayed and suppressed. In practice, the buildup of the LW background drives an increase in the minimum mass for star formation. 

In order to get a minimum mass scale for a halo to be able to efficiently cool in the presence of a photo-dissociating LW background, we compare two $\molh$ number density thresholds: the collapse threshold, which is set by the cooling time, $n_{H_2}^\mathrm{crit}$ (eq.~\ref{eq:nH2_crit}), and the dissociation equilibrium threshold, $n_{H_2}^\mathrm{eq}$ (i.e., the equilibrium $\molh$ density that results from a fixed LW background intensity). The former of these can be parameterized by the virial temperature of the halo, $T_\mathrm{vir}$, and the latter by the intensity in the LW bands, which is conventionally expressed in units of $J_{21} = 10^{-21} \mathrm{erg}\ \mathrm{s}^{-1}\ \mathrm{cm}^{-2}\ \mathrm{Hz}^{-1}\ \mathrm{sr}^{-1}$. Equating the two thresholds yields a maximum value of the LW intensity that allows a cloud with a fixed virial temperature to collapse or, equivalently, yields the critical virial temperature needed for collapse under a fixed value of the LW background intensity. 

The dissociation equilibrium threshold is set by  balancing the rate of $\molh$ formation and destruction, comparing the rates of photodissociation (pd) and collisional dissociation (cd):
\begin{equation}\label{eq:H2_eq_frac}
    f_{H_2}^\mathrm{eq} = \min\bigg(\frac{k_\mathrm{form}}{k_\mathrm{pd}}f_{e}, \frac{k_\mathrm{form}}{k_\mathrm{cd}}f_{e}\bigg)
\end{equation}
where we use the recombination code \texttt{CosmoREC} \citep{cosmorec} to compute the electron fraction $f_e$ and the reaction rate coefficients are the same as defined above \citep{SBH10}.

This equilibrium is modified by the effects of self-shielding---in halos where a sufficiently high column density of $\molh$ builds up, the gas can become optically thick to LW radiation and `shield' itself against photodissociation. This is usually parameterized in the form of a ‘shielding factor’, $f_{\rm shield}$, which reduces the photodissociation rate:
\begin{equation}
    k_\mathrm{pd}(N_{H_2}, J_{\rm LW}) = f_\mathrm{shield}(N_{H_2}, T)k_\mathrm{pd}(N_{H_2}=0, J_{\rm LW})
\end{equation}
so a value of $f_{\rm shield} = 1$ (0) would correspond to no (complete) shielding. We approximately calculate the column density using the peak halo gas density given in eq.~\ref{eq:ngas}---i.e., $N_{\molh} = f_{\molh} \lambda_\mathrm{J}n_\mathrm{\rm H}/2$, where $\lambda_J$ is the Jeans length in the halo gas cloud. We compute $f_\mathrm{shield}$ following \citet{WG11} (WG11) and \citet{WGH19} (WG19). Noting that the \citet{WGH19} expression is a correction to \citet{WG11} at higher densities, our shield factor is given by 
\begin{equation}\label{eq:fshield}
    f_\mathrm{shield} = \left\{\begin{array}{ll}
      f(\alpha = 1.1)\ (\mathrm{WG11}) & n\leq 10^{3}\ \mathrm{cm}^{-3} \\
      f(\alpha = \mathrm{eq\ 8})\ (\mathrm{WG19}) & \begin{aligned}[t]
      10^{3}\leq n/\mathrm{cm}^{-3} \leq 10^{7},\\  
      N_{H_2}\leq 10^{17}\ \mathrm{cm}^{-2}, \\
      T \leq 8000\ \mathrm{K}
      \end{aligned}
      \\
      1 & \mathrm{otherwise} \\
\end{array} 
\right.
\end{equation}
where the limits reflect those set by WG11/19.
Equating eq.~\ref{eq:nH2_crit} and $n_\mathrm{eq} = n_{\rm H} f_{\molh}^\mathrm{eq}$ (given by eqs.~\ref{eq:ngas} and~\ref{eq:H2_eq_frac}), we numerically solve for the virial temperature necessary for a gas cloud to collapse in the presence of a fixed LW background intensity. In the presence of a strong LW background, halos may be in the high mass limit of eq.~\ref{eq:ngas}, so we fit those independently and summarize the associated minimum masses as follows. For the low-mass limit, we find
\begin{equation}\label{eq:LW_low}
    \begin{split}
    M_{\rm LW, low} \simeq\ & 3.35\times 10^4 \bigg(\frac{\zeta}{0.25}\bigg)^{-0.19} \\ &\times \Bigg[1+13\bigg(\frac{J_{\rm LW}}{J_{21}}\bigg)^{0.38}\Bigg]\bigg(\frac{1+z}{21}\bigg)^{-1}M_\odot
    \end{split} \\
\end{equation}
At high masses, we find
\begin{equation}\label{eq:LW_high}
    \begin{split}
    M_{\rm LW, high} \simeq\ & 1.05\times 10^4 \bigg(\frac{\zeta}{0.25}\bigg)^{-0.33} \\ &\times \Bigg[1+26\bigg(\frac{J_{\rm LW}}{J_{21}}\bigg)^{0.62}\Bigg]\bigg(\frac{1+z}{21}\bigg)^{-3.91}M_\odot
    \end{split} \\
\end{equation}
Noting that the transition for these two regimes in the halo density occurs at $M_{\rm turn}$ (as defined in eq.~\ref{eq:M_turn}), the photodissociation mass is set by
\begin{equation}\label{eq:LW_threshold}
    \begin{split}
    M_{\rm LW} = \max\Big[\min\big(&M_{\rm LW, low}, M_{\rm turn}\big), \\
    & M_{\rm LW, high}\mathcal{H}\big(M_{\rm LW,high}-M_{\rm turn}\big)\Big]
    \end{split}
\end{equation}
where $\mathcal{H}(t-t_0)$ is the Heaviside step function (which returns 0 for $t < t_0$ and 1 for $t\geq t_0$). At a fixed redshift as a function of increasing $J_{\rm LW}$, this means we select the low-density result until that crosses the turnover mass, at which point we select the larger of the turnover mass or the high-density result.

\begin{figure}
    \centering
    \includegraphics[width = \columnwidth]{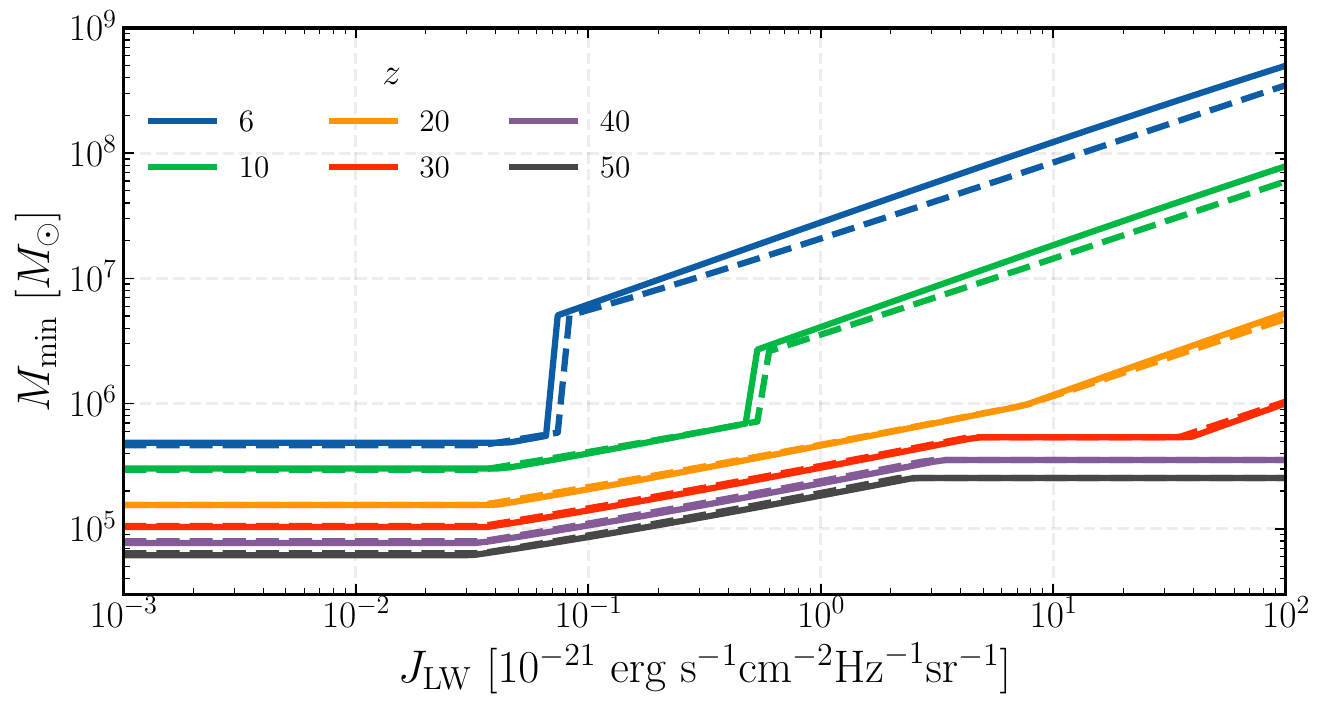}
    \caption{The minimum mass for a halo to form $\molh$ in the presence of a Lyman-Werner background, calculated for a range of LW backgrounds following the steps described in Sections~\ref{sssec:cooling_threshold} and~\ref{sssec:LW_bckd} at various redshifts (colored lines). The fitting formulae for the cooling and dissociation thresholds (eqs.~\ref{eq:cooling_threshold} and~\ref{eq:LW_threshold}, respectively) are overlaid with dashed lines.}
    \label{fig:mcrit_jlw}
\end{figure}

The results of this calculation for various redshifts are given in Figure~\ref{fig:mcrit_jlw}. Because this portion of the calculation does not account for the buildup of $\molh$ (eq.~\ref{eq:molH_frac}), we must still also compare this to the cooling threshold presented in \S~\ref{sssec:cooling_threshold} and check that a halo has enough $\molh$ to efficiently cool (i.e., we choose the larger of eqs.~\ref{eq:cooling_threshold} and~\ref{eq:LW_threshold}). The cooling threshold sets the baseline for the minimum mass for small values of the LW intensity, with the turnover decreasing with redshift. Once the LW background takes over, however, the critical mass increases with increasing LW background intensity at a fixed $z$.

We note that the amplitude of the LW background will be computed self-consistently by our semi-analytic model, as described in \S~\ref{sec:semianalytic_model}.

\subsection{Thermalization threshold}\label{sssec:thermalization_threshold}

The cooling argument of \S~\ref{sssec:cooling_threshold} essentially requires that the halo produce enough molecular hydrogen in order for its radiative cooling to shed the thermal energy generated with gravitational infall. However, in the presence of streaming, it must also shed the excess kinetic energy carried by the baryons (which will also be thermalized during collapse). This effect can be described as follows (\citealp{Fialkov12}): 
\begin{equation}\label{eq:vbc_alpha}
    T_{\rm vir} \mapsto T_{\rm vir} + \frac{\mu m_p (\alpha_{\rm vbc} v_{\rm bc})^2}{2k_B}
\end{equation}
where $\alpha_{\rm vbc}$ parameterizes the magnification of the stream velocity as gas collapses into the halo potential well and has been shown to take on values of $\sim 4-6$ in simulations (\citealp{Fialkov12}, \citealp{McQuinnOleary12}). 

We can understand the numerical value of $\alpha_{\rm vbc}$ by the following two arguments \citep{McQuinnOleary12}. First, we  consider the case of accreting shock heated gas as it collapses into the halo. In adiabatic collapse, $\rho\propto T^{3/2}$, so the overdensity of the virialized gas with no stream velocity is 
\begin{equation}\label{eq:shock_heat}
    \delta_b \sim \bigg(\frac{T_{\rm vir}}{T_{\rm IGM}^{\rm ad}}\bigg)^{3/2}
\end{equation}
 The thermal energy associated with the stream velocity provides an extra contribution to the IGM temperature, corresponding to $T_{\rm IGM}\mapsto T_{\rm IGM}(1+5\mathcal{M}_{\rm bc}^2/9)$, where $\mathcal{M}_{\rm bc} = v_{\rm bc}/c_{\rm s, IGM}$ is the mach number of the streaming gas.\footnote{This can be seen if we assume that the energy in the stream velocity is converted into thermal energy in the IGM. That is, the kinetic energy associated with streaming gives $\mu m v_{\rm bc}^2/2 = \mu m c_s^2 \mathcal{M}_{\rm bc}^2/2$. Using the expression for the sound speed of a monatomic ideal gas and assuming some fraction $\beta$ of this energy is converted into thermal energy (as seen in the simulations of \citealp{OlearyMcquinn12} and \citealp{McQuinnOleary12}), then the temperature is increased from $T\to T'$ by a factor of $f \equiv T'/T = 1+5\beta\mathcal{M}_{\rm bc}^2/9$. We have taken $\beta = 1$ to compute the ballpark value above.} With eq.~\ref{eq:shock_heat}, this means that the gas density in the halo scales as $n_{\rm H}\propto \delta_b \propto (1+5\mathcal{M}_{\rm bc}^2/9)^{-3/2}$. We can fold this additional contribution to the gas density into equation~\ref{eq:nH2_crit} and calculate the new virial temperature necessary for efficient cooling with the same argument as that outlined in \S~\ref{sssec:cooling_threshold}. Combining this result with equation~\ref{eq:vbc_alpha} suggests $\alpha_{\rm vbc}\sim 6$ at $z=20$. However, as the critical $\molh$ fraction, IGM temperature, and stream velocity all evolve with redshift, the value of $\alpha_{\rm vbc}$ given by this calculation will as well.

Next, we can consider that velocities will be boosted by a factor of $(1+\delta)^{1/3}$ in an adiabatically collapsing region, so the effective stream velocity will be magnified to $\approx 6v_{\rm bc}$ for overdensities of $\delta\sim 200$. If the circular velocity of the halo is to be larger than this local stream velocity, the circular velocity necessary for the high velocity gas to be captured by the halo potential well is
\begin{equation}
    v_{\rm c}(v_{\rm bc}) = \sqrt{v_{\rm c}^2 + (6v_{\rm bc})^2/3}
\end{equation}
where we divide by a factor of 3 because the local stream velocity will have a particular directionality, whereas the halo gas moves isotropically after thermalization. Converting this velocity to the associated virial temperature corresponds to a redshift-independent value of $\alpha\sim 4$.

These analytic arguments give us a ballpark value of $\alpha_{\rm vbc} \sim \mathcal{O}(1-10)$ on average. For our fiducial calculations, however, we defer to the more detailed numerical simulations of \citet{McQuinnOleary12} (who find $\alpha_{\rm vbc}\sim 4-6$) and therefore set $\alpha_{\rm vbc} = 5$.

In practice, we incorporate this into our model by adding this extra contribution to the virial temperature (eq.~\ref{eq:vbc_alpha}) to the result of the calculation in Sections~\ref{sssec:LW_bckd} and~\ref{sssec:cooling_threshold}. This contribution to the mass scale can be parameterized as a modification to the virial velocity of the halo. That is, for our chosen cosmology, the virial velocity of a halo without the effects of streaming is given by \citep{BL01}
\begin{equation}\label{eq:v_vir_nostream}
    v_0 = 5.28\times 10^{-2} M_0^{1/3}\bigg(\frac{1+z}{21}\bigg)^{1/2}\ {\rm km\ s^{-1}}
\end{equation}
where $M_0$ is the larger of eqs.~\ref{eq:cooling_threshold} and~\ref{eq:LW_threshold}. Then, the thermalization threshold is given by
\begin{equation}\label{eq:thermalization_threshold}
    M_{\rm bc, LW, cool} \simeq 6.79\times 10^3 \Big[v_0^2 + (\alpha_{\rm vbc}v_{\rm bc})^2\Big]^{3/2} \bigg(\frac{1+z}{21}\bigg)^{-3/2} M_\odot
\end{equation}

\subsection{X-ray background}\label{sssec:xray-bkgd}
Unlike the Lyman-Werner dissociation and stream velocity, the buildup of a metagalactic X-ray background both negatively \textit{and} positively feeds back into the minimum mass (and thus the star formation rate). A background of X-ray radiation can photoionize the IGM, boosting the electron fraction and heating the gas. The increased density of free electrons can then promote the formation of $\molh$, supporting star formation, while the heating can make accretion of gas more difficult. Some energy from this radiation goes into ionizing the hydrogen and helium atoms and the rest goes to the electron, heating it. Following the discussion in \citet{Furlanetto06}, the hot electron distributes its energy by collisionally ionizing other atoms, producing secondary electrons, collisionally exciting H and He, and undergoing Coulomb collisions with thermal electrons. The cross sections of these processes set the fraction of the energy that goes into heating and the fraction that goes into ionization, for which we use the expressions of \citet{ShullvanSteenberg85}: 
\begin{align}
    f_\mathrm{X,h} &= C_1\Big[1-(1-x_e^{a_1})^{b_1}\Big] \\
f_{\rm X, ion} &= C_2(1-x_e^{a_2})^{b_2}
\end{align}
Note that we test the effects of more accurately including secondary ionizations (see e.g., \citealp{Furlanetto10}, \citealp{Ricotti16}) and find that they do not meaningfully change the results.

Therefore, calculating the effects of an X-ray background requires a calculation of the IGM temperature and electron fraction. As described in \citet{Furlanetto06}, the X-ray contribution to the temperature and electron fraction of the IGM can be written in terms of the X-ray emissivity $\epsilon_X$ as follows:
\begin{align}
    \frac{dT_X}{dz} &= \frac{2}{3}f_{\rm X, h}\frac{\epsilon_X}{k_B n}\frac{dT_X}{dt}\frac{dt}{dz} = -\frac{2}{3}\left(\frac{f_{\rm X,h}}{0.2}\right)\frac{\epsilon_X}{k_B n H(z) (1+z)} \\
    \frac{dx_e}{dz} &= -f_{\rm X, ion}\frac{\epsilon_X}{E_H n H(z) (1+z)}\label{eq:xe_ev}
\end{align}
where $E_H = 13.6\ {\rm eV}$ is the ionization energy of hydrogen. We calculate a spectrum-averaged X-ray emissivity from the $L_X-{\rm SFR}$ relation, which we scale by a factor $f_X$ to account for deviations from the local calibration (e.g., \citealp{Mineo12}, \citealp{Lehmer16})\footnote{Theoretical models (e.g., \citealp{Mesinger13}, \citealp{Fragos13}) predict that the $L_X-{\rm SFR}$ relation will increase with redshift, with $f_X\sim 10-50$ at $z\sim 10$ suggested by the results of \citet{Greig21} and \citet{Hera22}. An upper limit on the value of $f_X$ is placed by the \textit{Chandra Deep Field}-South observations of the unresolved soft X-ray background (SXB; \citealp{Hickox07}); $f_X \gtrsim 100$ would saturate the SXB \citep{Mcquinn12}, so we bound our models by testing values between $f_X\sim 1-100$.}:
\begin{equation}\label{eq:SFR_emissivitty}
\epsilon_{X}(z) =2.6 \times 10^{39} f_X \left(\frac{\rho_{\rm SFR}(z)}{M_{\odot} \mathrm{yr}^{-1} \mathrm{cMpc}^{-3}}\right) \mathrm{erg}\ \mathrm{s}^{-1} \mathrm{cMpc}^{-3}
\end{equation}
where $\rho_{\rm SFR}$ is the star formation rate density computed from our model. 

\begin{figure}
    \centering
    \includegraphics[width=\columnwidth]{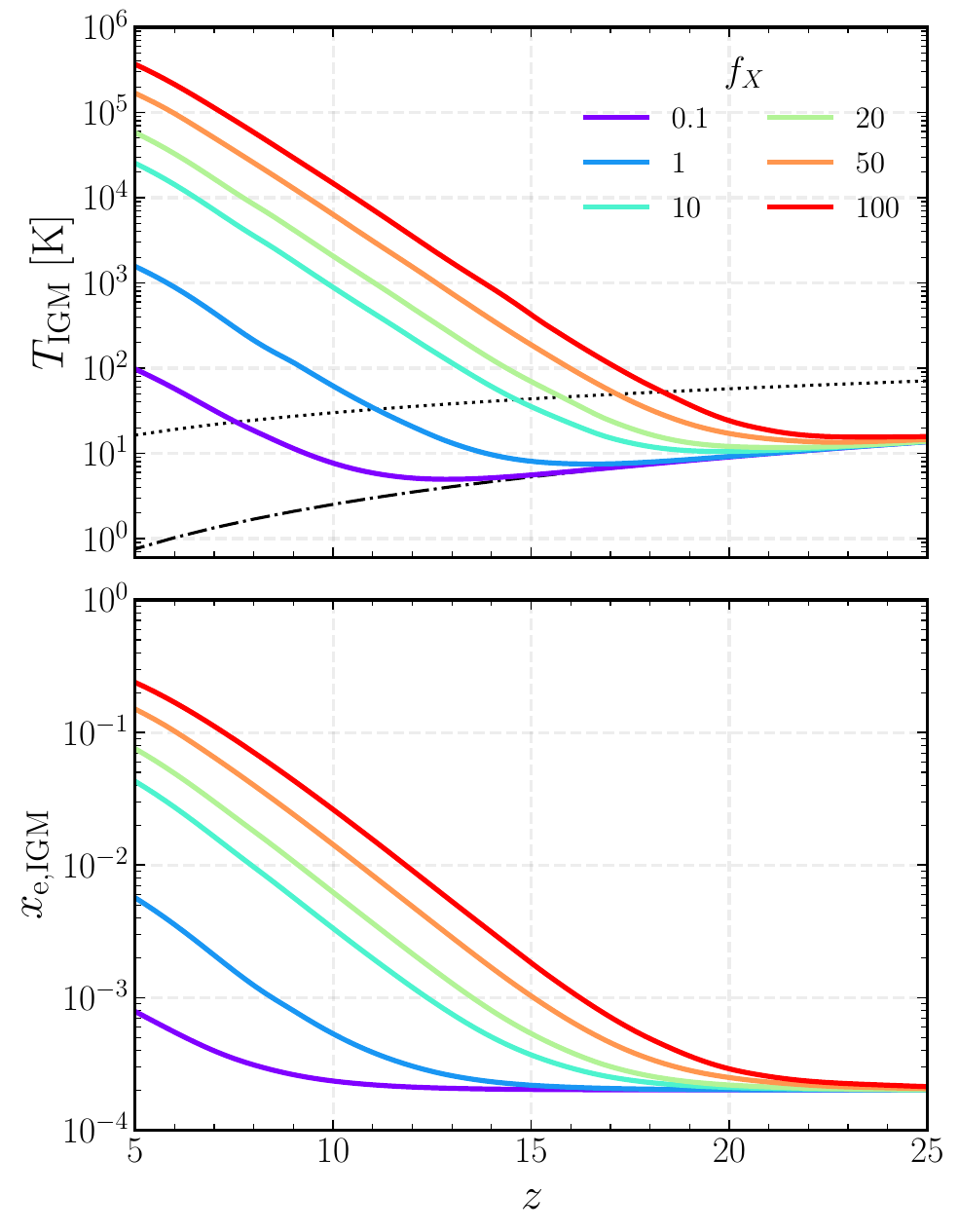}
    \caption{The evolution of the IGM temperature (top) and electron fraction (bottom) with redshift for various values of $f_X$ using the star formation history computed with the fiducial parameters of \citet{Mebane18}. We compare our calculation to that of \texttt{CosmoREC} (i.e., no X-ray heating) with a black dot-dashed curve and show the CMB temperature with a black dotted curve.}
    \label{fig:xray_heat_ion}
\end{figure}

Initializing the calculation with the IGM temperature and electron fraction calculated with \texttt{CosmoREC}, we integrate the full evolution equations (given in \citealp{Furlanetto06}) simultaneously, yielding the histories displayed in Figure~\ref{fig:xray_heat_ion}.

With these in hand, we compute the X-ray contribution to the minimum mass through its effects on the previously discussed quantities: the filter mass and LW mass. To incorporate the effects of heating, we modify the temperature evolution of the IGM used in the calculation of the filter mass (see \S~\ref{sssec:filter_mass}). This will have the effect of raising the filter mass at late times as the IGM temperature rises steeply for $z\lesssim 15$ (Figure~\ref{fig:xray_heat_ion}). Ionization raises the free electron fraction of the IGM. The free electrons catalyze the formation of $\molh$, hastening collapse and lowering the minimum mass scale derived from LW photodissociation (see \S~\ref{sssec:LW_bckd}). 

However, because of the higher density and thus recombination rate within a halo, the relevant electron fraction within a DM halo will be smaller than the IGM value that we calculate from integrating equation~\ref{eq:xe_ev}. We can estimate the electron fraction in a halo by appealing to a similar argument as that outlined in \S~\ref{sssec:cooling_threshold}. That is, if we assume that $n_{\rm HII} = n_{\rm e}$, then eq.~\ref{eq:H_rate_eqns} will describe the evolution of $x_{\rm e}$ as well. This then admits the same solution as before; that is:
\begin{equation}\label{eq:xe_rec}
    x_{\rm e, rec} = \frac{x_{\rm e}^{\rm IGM}}{1 + t_H/t_{\rm rec}^i}
\end{equation}
where $t_{\rm rec}$ is defined as before and we have taken $t=t_H$ to be the relevant timescale for comparison here.

Therefore, there are two relevant limits for this calculation. If the recombination time is long compared to the Hubble time (so $t_{\rm rec}\gg t_H$), then the cloud will not reach equilibrium and thus the electron fraction in the halo will be given by eq.~\ref{eq:xe_rec}. If the recombination time is sufficiently short compared to the Hubble time (i.e., $t_{\rm rec}\ll t_H$), then the solution (eq.~\ref{eq:xe_rec}) no longer applies---because the ionizing background could raise the electron fraction during that time---and we must instead calculate the equilibrium level of ionization. 

For the equilibrium calculation, we balance ionization and recombination, solving
\begin{equation}
    \alpha_B n_e n_p = \Gamma_X n_{\rm HI}
\end{equation}
where $\Gamma_X$ is the ionization rate. Because we do not directly model the X-ray source spectrum (i.e., $\epsilon_X$ is taken to be frequency-independent), the ionization rate can be found by solving the cosmological radiative transfer equation \citep{HaardtMadau12}:
\begin{equation}\label{eq:ion_rate}
    \Gamma_X(z) = c\int_{\nu_{\rm thresh}}^\infty d\nu \frac{\sigma_{\rm HI}(\nu)}{h\nu}\int_z^\infty dz'\frac{(1+z)^3}{H(z')(1+z')^4}\epsilon_{X}(z')e^{-\tau}
\end{equation}
where the threshold frequency $h\nu_{\rm thresh} = 13.6\ {\rm eV}$, $\nu' = \nu(1+z')/(1+z)$, and the optical depth is $\tau = \int_z^{z'} (d\ell / dz'') dz''[n_{\rm HI}\sigma_{\rm HI}(\nu'') + n_{\rm HeI}\sigma_{\rm HeI}(\nu'')]$. For the cross sections $\sigma_{\rm HI}$ and $\sigma_{\rm HeI}$, we use the fits of \citet{Verner96}. 

As we evolve the X-ray background throughout our calculation, we fold the newfound halo electron fraction into our calculation of the LW mass (in eq.~\ref{eq:H2_eq_frac}), and expect that it will suppress the mass scale as $x_{\rm e, halo}$ increases. 

Including the effects of X-rays in a simple fitting formula is comparatively more difficult than the preceding effects. In particular, to incorporate the effects of heating, one must integrate over the full temperature history to accurately estimate the filter mass. Therefore, we only present a simple modification to the fitting formulae for the photodissociation threshold to account for the effect of X-ray ionization:
\begin{align}\label{eq:ionization_threshold}
    M_{\rm LW, Xray, low} &\simeq M_{\rm LW, low}\bigg(\frac{x_{\rm e, halo}}{x_{\rm e, no\ Xray}}\bigg)^{-0.19} \\
    M_{\rm LW, Xray, high} &\simeq M_{\rm LW, high}\bigg(\frac{x_{\rm e, halo}}{x_{\rm e, no\ Xray}}\bigg)^{-0.33}
\end{align}
where $M_{\rm LW,low}$ and $M_{\rm LW, high}$ are given by equations~\ref{eq:LW_low} and~\ref{eq:LW_high}, and $x_{\rm e, no\ Xray}$ is the electron fraction of the IGM without the effects of an X-ray background.\footnote{For this calculation, one can refer to a recombination code such as \texttt{RECFAST} or \texttt{CosmoREC} (\citealp{recfast}, \citealp{cosmorec}). For the relevant redshifts for Pop III star formation (i.e., between $z=5$ and $50$), however, the electron fraction without X-rays is well fit by a power-law of the form $x_{\rm e, no\ Xray} \approx 2.19\times 10^{-4} [(1+z)/21]^{0.12}$.} From these formulae, we can see that an increase in the electron fraction from the IGM electron fraction without the effects of an X-ray background will indeed suppress the dissociation mass, though the strength of this contribution will depend on the density regime we are in.

\subsection{Summary}\label{sssec:fitting_formula}
Our minimum mass model can be summarized as follows. A halo must first exceed the filter mass threshold (\S~\ref{sssec:filter_mass}; eq.~\ref{eq:filter_threshold}) in order to be able to accrete baryons. Once this gas is accreted, a sufficiently large fraction of $\molh$ must build up such that the gas can efficiently cool and collapse to high densities (\S~\ref{sssec:cooling_threshold}; eq.~\ref{eq:cooling_threshold}). If the halo is in a region of the universe with a large DM-baryon relative velocity, the halo must also overcome the additional thermal energy associated with the relative motion (\S~\ref{sssec:thermalization_threshold}; eq.~\ref{eq:thermalization_threshold}). As the first generations of stars form, they produce a photodissociating LW background (parameterized in terms of the specific intensity $J_{\rm LW}$) and a photo-heating and -ionizing X-ray background (which increases the temperature of the IGM and the electron fraction of the halo $x_{\rm e, halo}$). Strong LW intensities 
and efficient X-ray heating can significantly boost the minimum mass (\S~\ref{sssec:LW_bckd} and~\ref{sssec:xray-bkgd}; eq.~\ref{eq:LW_threshold}), while ionization can promote the formation of $\molh$ and lower the mass (eq.~\ref{eq:ionization_threshold}).

In practice, this means the minimum mass is given by 
\begin{equation}\label{eq:overall_threshold}
    M_{\rm min} = \max[M_F, M_{\rm bc, LW, cool}]
\end{equation}
where $M_F$ is given by eq.~\ref{eq:filter_threshold} and $M_{\rm bc, LW, cool}$ is given by eq.~\ref{eq:thermalization_threshold} (which depends on eqs.~\ref{eq:cooling_threshold},~\ref{eq:LW_threshold}, and~\ref{eq:ionization_threshold}). These fitting formulae presented in the preceding several sections are accurate to within 40\% of the full calculation, and we find that using them does not change the results of the semi-analytic calculation (if we do not include the effects of IGM heating). IGM heating appears to be important at the latest times and in the case of the strongest X-rays; that is, when $f_X \gtrsim 50-100$. This contribution to the minimum mass is discussed in more detail in \S~\ref{ssec:min_mass_variations}. 

Calculating the minimum mass and understanding how and when individual components of the physics are important requires robust characterization of the star formation histories and associated radiation backgrounds. As such, we now explore the effects of this minimum mass model in the context of a semi-analytic model for high-redshift star formation.

\section{Semi-analytic Model}\label{sec:semianalytic_model}
We base our calculations on an updated version of the semi-analytic model presented in \citet{Mebane18} (hereafter, M18). This model employs a feedback-limited star-formation prescription, wherein star formation is tracked in individual halos until feedback shuts off the formation process. We follow a sample of 100 halos from $z=50$ to 6 with masses ranging from $10^{6}-10^{13} M_\odot$ at $z=6$ with timesteps of 1 Myr. The growth history of these halos is calculated using abundance matching, where we assume that halos maintain their comoving number density through cosmic time, with the halo mass function given in \citet{Trac15}. Star formation is followed in individual halos by comparing the halo mass to the evolving minimum mass outlined in \S~\ref{sec:min_mass}. Once a halo passes this star formation threshold, we compare its gas mass (which is given by $M_{\rm gas} = \Omega_b/\Omega_m M_{\rm h}$ until supernovae evacuate some fraction of this gas) to the local Jeans mass. If the gas mass exceeds the local Jeans mass, it will randomly form one or two stars (determined by a fixed binary probability $f_{\rm bin} = 0.5$) with masses drawn from a Chabrier-like IMF \citep{Chabrier03}:
\begin{equation}\label{eq:chabrier}
    \frac{dN}{dm} \propto M^{-\alpha}\exp\Bigg[-\bigg(\frac{M_{\rm char}}{M}\bigg)^\beta\Bigg]
\end{equation}
where $\alpha$ is the Salpeter-like power-law slope that characterizes the high-mass end of the IMF and $M_{\rm char}$ is the characteristic mass. These massive stars live for 5 Myr and their end-of-life behavior is determined by their stellar mass; i.e., stars with $M/M_\odot \in [8,40]\cup[140,260]$ end their lives in SNe and the rest collapse directly to BHs.  

Because Pop III stars form in minihalos at high redshifts, supernova feedback can significantly affect their environments through metal enrichment and gas ejection. When a supernova occurs in a halo, the circumstellar gas is enriched according to the metal yields given by \citet{HegerWoosley02} and \cite{HegerWoosley10} for pair-instability and core-collapse SNe, respectively. We then assume that 10\% of the released kinetic energy couples to gas in the halo and ejects it. The mass and metallicity of the ejected gas is then followed and allowed to reaccrete after a free-fall time (roughly 50 Myr for a $10^5M_\odot$ minihalo at $z = 15$; see \S 6.4 in \citealp{Mebane18} for a detailed discussion of this choice). In the meantime, the halo continues to grow via accretion from the pristine IGM, enabling the potential formation of subsequent generations of Pop III stars. We note that the ability for halos to form multiple generations of Pop III stars is a unique feature of our model that we discuss in more detail in \S~\ref{ssec:sam_comparison}. 
The simulations of \citet{Abe21}---which modeled the effects of SN feedback and metal enrichment on Pop III star formation---found that the delay in reaccretion of enriched gas allowed for multiple generations of Pop III star formation before a halo crossed the critical metallicity and transitioned to the Pop II regime, which motivates our choice here.

The transition to Pop II star formation is an important piece of our model, as Pop II stars will dominate in the buildup of the metagalactic LW and X-ray backgrounds. A halo will transition to Pop II star formation once it has reached the atomic cooling threshold ($T_{\rm vir} > 10^4$ K) or has been sufficiently enriched (i.e., the mean metallicity of the halo exceeds the critical CII and OI concentrations given in \citealp{BrommLoeb03}). We note that we assume that atomic-cooling haloes form Pop II stars without attempting to model metal mixing in those haloes, but even if metal mixing is inefficient, their rapid radiative cooling will modify the process of star formation so that $\molh$ is no longer required. Any Pop III stars forming in massive haloes are thus likely to be more similar to Pop II stars than those forming inside minihaloes.

Once a halo transitions, we use the bursty star formation prescription outlined in \citet{Furlanetto22_burst}. This  modifies the feedback-regulated models of \citet{Furlanetto17} to include a feedback delay that accounts for the nonzero lifetimes of SN progenitors. That is, rather than instantaneously injecting SN feedback into the system after stars form, we include a 5-30 Myr delay (accounting for stellar lifetimes) that allows for continued star formation in the meantime. As a result, the star formation rate can `overshoot' the expectation of a simple feedback-limited model and halos will ultimately go through cycles of star formation in characteristic `bursts.' This effect is most pronounced in the least massive halos, which are the most susceptible to the effects of feedback. As a halo grows, these star formation cycles damp out and the star formation efficiency approaches the equilibrium result used in M18 (originally presented in \citealp{Furlanetto17}).

We use the aforementioned framework to evolve the system of halos between redshifts 50 and 6. We extrapolate these results (which are calculated for a sample of 100 halos) to calculate global quantities, such as the star-formation rate density (SFRD), by averaging over the halo mass function. To minimize noise in our calculations (because we track only 100 halos for computational efficiency), at every timestep, we inject 10,000 `fake' halos into our calculation. These halos form stars with a probability given by the star formation duty cycle $f_{\rm duty}(z) = N_{\rm III, on}(z)/N_{\rm III}(z)$, where $N_{\rm III, on}(z)$ is the number of halos that are actively forming stars and $N_{\rm III}(z)$ is the number of halos that are able to form Pop III stars at a given $z$ (i.e., have yet to transition to Pop II star formation). We compare this procedure to calculating the SFRD with 1,000 `real' halos (compared to the 100 in our fiducial runs) and find that the `fake' halo injection robustly reproduces the expected SFRD.  

\begin{figure*}
    \centering
    \includegraphics[width=\textwidth] {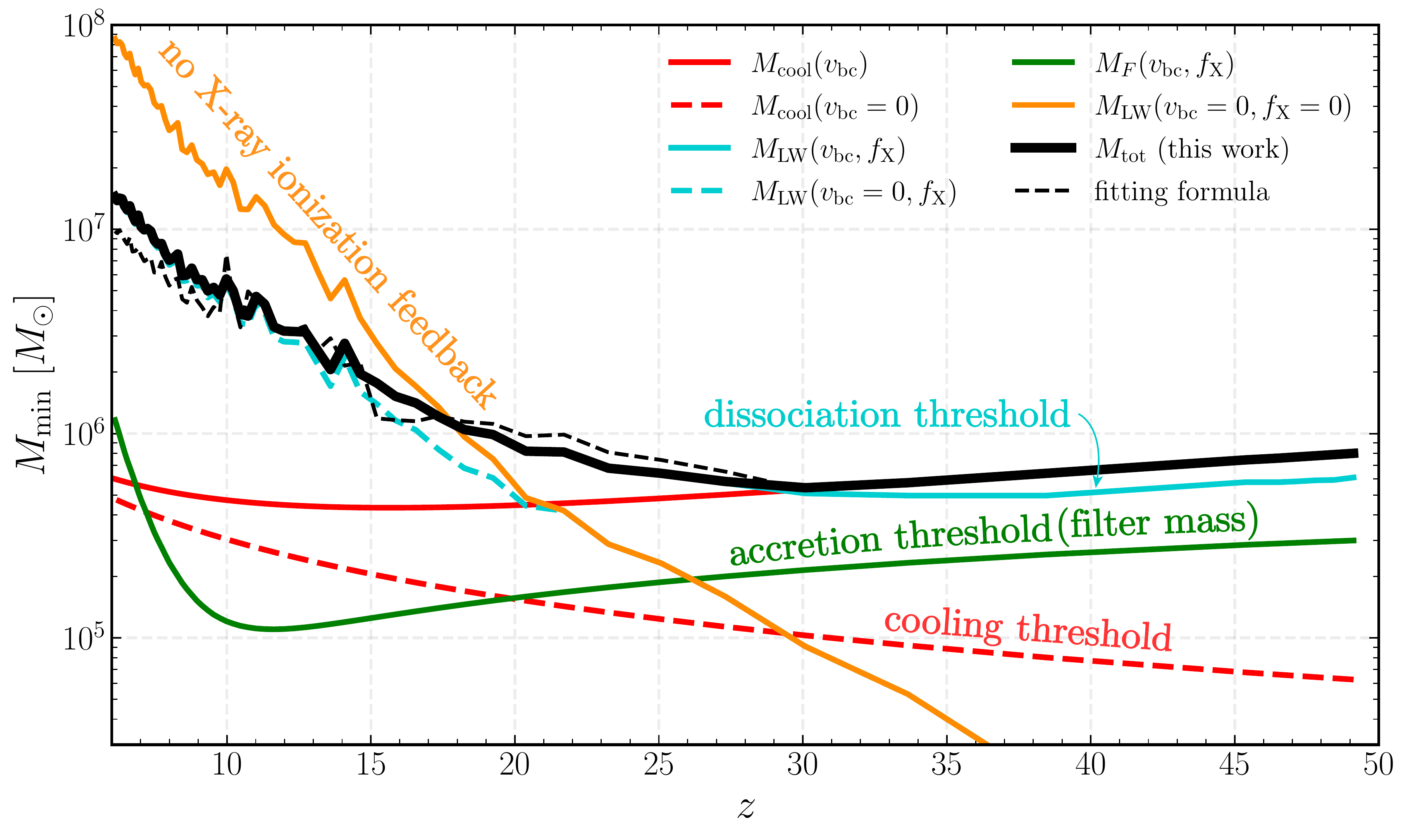}
   
    \caption{The minimum DM halo mass for Pop III star formation as a function of redshift in our fiducial model, which has $v_{\rm bc} = \sigma_{\rm vbc}$ and $f_X = 10$ (solid  black curve) and the associated fitting formula (dashed black curve). We present the individual components of the calculation following \S~\ref{sec:min_mass}: the cooling threshold (red), the accretion threshold (green), and the photodissociation threshold (blue). For the cooling and dissociation thresholds, the equivalent calculation with $v_{\rm bc} = 0$ is given as a dashed curve of the same color. The dissociation threshold without the calculation of the X-ray background (\S~\ref{sssec:xray-bkgd}) is given in orange.}
    \label{fig:min_mass}
\end{figure*}

\section{Results}\label{sec:results}

We now use the updated prescription for the minimum mass (\S~\ref{sec:min_mass}) and the improvements to the semi-analytic framework (\S~\ref{sec:semianalytic_model}) to follow star formation during the Pop III era.
For the fiducial model:
\begin{enumerate}[(i)]
    \item In calculating the minimum mass, we assume a stream velocity magnitude of $v_{\rm bc} = 1\sigma_{\rm vbc}$ and an ${\rm L}_X-{\rm SFR}$ scaling of $f_X = 10$.
    \item We assume a Chabrier IMF (equation~\ref{eq:chabrier}) with a maximum mass of $500 M_\odot$, $\alpha = 2.35$ \citep{Salpeter55}, $\beta = 1.6$, and $M_{\rm char} = 20 M_\odot$. 
    \item We include bursty star formation for Pop II halos.
\end{enumerate}
We note that our fiducial value for the magnitude of the stream velocity ($1\sigma_{\rm vbc}$) was chosen for ease of comparison with existing work. The stream velocity follows a Maxwell-Boltzmann distribution, for which the mean is $\sqrt{8/3\pi}\sigma_{\rm vbc}\sim 0.92\sigma_{\rm vbc}$ and the most probable value is $\sqrt{2/3}\sigma_{\rm vbc}\sim 0.82\sigma_{\rm vbc}$. We find that choosing one of these values would lower the minimum mass at early times by a factor of a few but would be qualitatively similar to the $1\sigma_{\rm vbc}$ case. Our fiducial choice of $f_X = 10$ is motivated by the findings of \citet{Hera22} (and the predictions of \citealp{Fragos13}). We also present variations of many of these parameters around these fiducial values.

\begin{figure}
    \centering
   \includegraphics[width=\columnwidth]{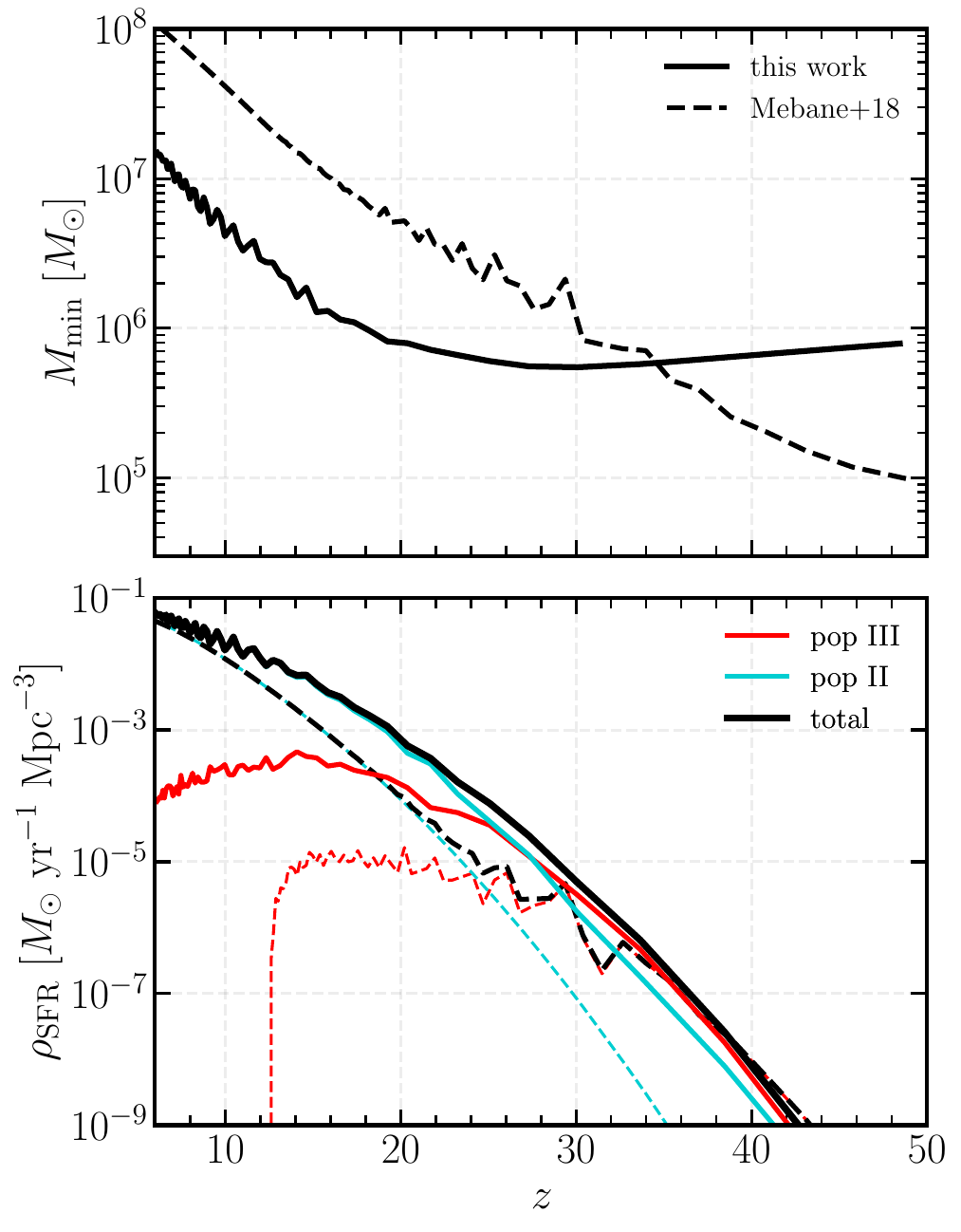}
    \caption{(top) The minimum DM halo mass for Pop III star formation as a function of redshift in our fiducial model ($v_{\rm bc} = 1\sigma_{\rm vbc}, f_X = 10$; solid black curve) and the previous incarnation of the semi-analytic model (M18; dashed black curve). (bottom) The star formation rate density (SFRD) as a function of redshift comparing our fiducial model (solid) to M18 (dashed). Note that we are using a different fiducial parameter set from M18 for this comparison. The Pop III (II) contribution to the SFRD is given in red (blue) and the total SFRD is given in black.
    }
    \label{fig:fid_SFRD}
\end{figure}

In Figure~\ref{fig:min_mass}, we present the minimum mass calculated using the radiation backgrounds computed with our fiducial parameter set (see lower panel of Figure~\ref{fig:fid_SFRD} for the associated SFRD). We can understand the evolution of the minimum mass with time by inspecting the evolution of the individual contributions to the minimum mass (colored curves) in various epochs. 

\begin{enumerate}[(I)]
    \item \textbf{Streaming dominance}, $z\gtrsim 30$: At the earliest times, the stream velocity is the dominant 
    limiting factor and the shape of the minimum mass traces the evolution of the cooling threshold for a region of 1$\sigma_{\rm vbc}$ streaming (red). The comparison between the cooling threshold with and without streaming (solid and dashed red curves, respectively) demonstrates the significant effect of the stream velocity on the minimum mass (nearly two orders of magnitude magnification). During this time, the 
    minimum mass decays with the stream velocity as $v_{\rm}\propto (1+z)$ and the calculations with and without streaming (dashed vs solid curves) rapidly converge.
    \item \textbf{LW dominance}, $10\lesssim z\lesssim 30$: By $z=30$, the SFRD has grown (largely dominated by Pop III stars; the solid red curve in the lower panel of Figure~\ref{fig:fid_SFRD}) and the LW background has built up, so the LW mass (blue) takes over and the minimum mass begins to grow. Coincident with this is an almost plateau in the Pop III SFRD, as the minimum mass is now growing 
at a rate comparable to the growth rate of low mass halos. This redshift evolution of the LW mass is tied to the efficiency of self-shielding in massive halos. In other words, because of the steep decrease in halo gas density as the universe expands
    (i.e., $n_{\rm gas}\propto (1+z)^{3}$), self-shielding is less efficient at lower $z$ and the LW mass can climb steadily as a result. 
    \item \textbf{Bursty star formation}, $7\lesssim z \lesssim 15$: At later times, Pop II star formation dominates and the effect of the bursty SF cycles is apparent in both the SFRD and the minimum mass. As a result, the Pop II SFRD overshoots the equilibrium case (black dashed curve in lower panel of Figure~\ref{fig:fid_SFRD}) and induces a slight increase in the minimum mass, as we expect from the results of \citet{Furlanetto22_burst}. We discuss this effect in more detail in \S~\ref{ssec:bursty_comparison}. 
    \item \textbf{X-ray era}, $z \lesssim 15$: At the latest times, we can compare the positive and negative effects of the X-ray background.  The dominant effect of the X-ray background is the positive feedback from photoionization, which suppresses the minimum mass more efficiently at lower $z$ (orange vs. black curve) as the universe becomes ionized (Figure~\ref{fig:xray_heat_ion}). With $f_X=10$ in our fiducical model, the negative effect of X-ray heating is subdominant and is manifested as an upturn in the filter mass at the latest times. Because of this minimum mass suppression (from ionization), Pop III star formation persists through the end of our calculation to $z\sim 6$. 
\end{enumerate}   

While the redshifts of transition between the different effects will vary with $v_{\rm bc}$ and $f_X$, this evolution of the minimum mass is qualitatively generic to all models, as is discussed below. 

\subsection{Comparison to M18}
Comparing our fiducial minimum mass to that of M18 during this epoch, we can isolate the key differences that result from the updated calculations described in \S~\ref{sec:min_mass}. The M18 model used a prescription for the minimum mass fit to the results of simulations \citep{OsheaNorman08} that did not include the effects of self-shielding. As the Lyman-Werner background builds up, we see a nearly order of magnitude suppression in the mass scale. The inclusion of self-shielding, following the updated criterion of WG11/WG19 (eq.~\ref{eq:fshield}), introduces more efficient shielding in the densest halos, suppressing the effects of the photodissociating background. Coupled with the positive effect of X-ray ionization,\footnote{The minimum mass prescription used in M18 does \textit{not} include the effects of an X-ray background --- without this positive feedback, the minimum mass would be increased by a factor of a few at late times; see \S~\ref{ssec:min_mass_variations}.} this has the effect of allowing a sustained and slowly growing level of Pop III star formation down to the latest times, when the SFRD begins to decline. Moreover, the peak Pop~III SFRD is nearly two orders of magnitude larger in the new model, largely because of the self-shielding in massive halos. As a result, Pop~III star formation will contribute significantly for a much longer time. The difference in the shapes of the minimum mass curves at $z\gtrsim 25$ is due to the inclusion of the stream velocity in the calculation of the cooling threshold. However, because of the additional requirement that the halo gas mass must exceed the local Jeans mass in order for star formation to begin, the nearly order of magnitude increase in the minimum mass only slightly filters into the global SFRD.

\begin{figure}
    \centering
    \includegraphics[width=\columnwidth]{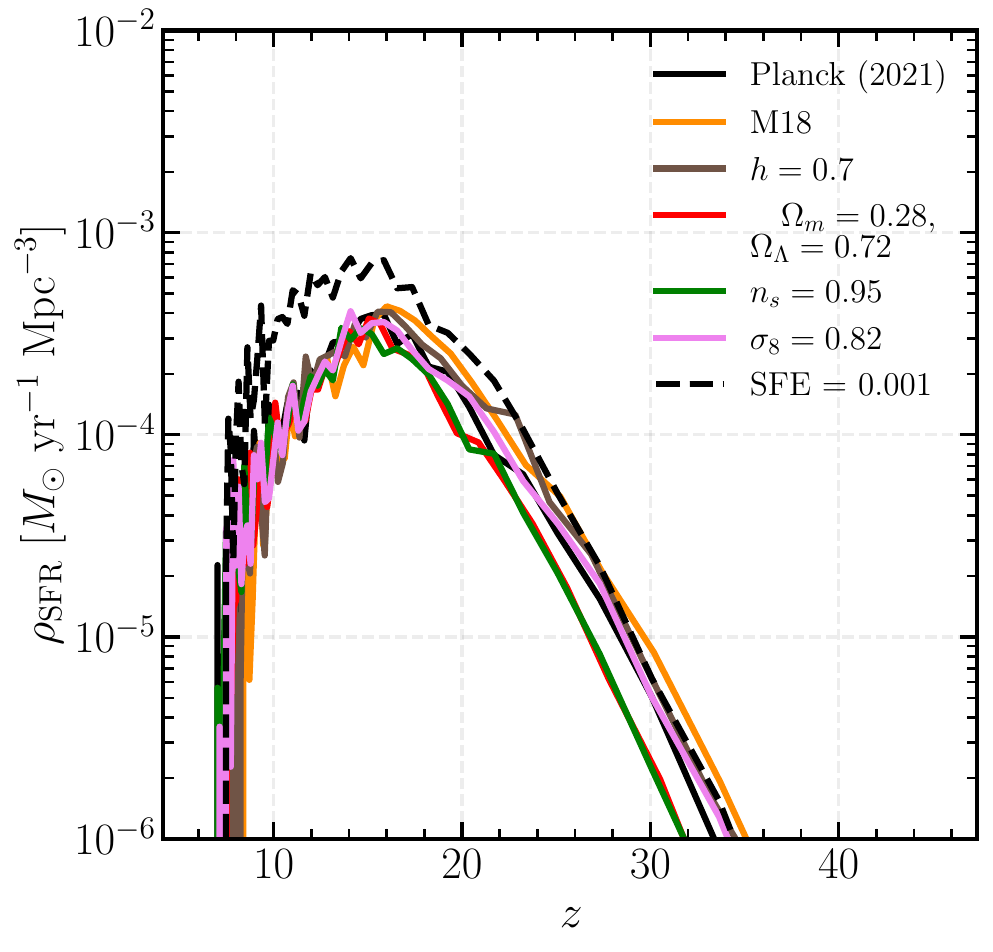}
    \caption{The redshift evolution of the Pop III SFRD for different choices of the underlying cosmology (different colors) and star-formation efficiency (SFE; dashed line) with $f_X, v_{\rm bc} = 0$. Our fiducial choice is given in black, the cosmology used in M18 in orange, and variations of one or a few parameters from our fiducial choice in the other colors.}
    \label{fig:SFRD_cosmo}
\end{figure}

In Figure~\ref{fig:SFRD_cosmo}, we compare the calculation of the Pop III SFRD using our fiducial cosmology to that of M18 and consider some individual parameter variations to diagnose the differences. Disentangling the sources of discrepancy induced by changes in the overall cosmology is challenging, but it is clear that the early phases of Pop III star formation are sensitive to this choice. Once the radiation backgrounds build up and the SFRD reaches its peak, however, the differences are washed out and the various models converge to the same result. 

Some simulations note that Pop III stars are likely to form in small groups or clusters (e.g., \citealp{Greif11}, \citealp{Latif22}). In our fiducial model we assume that Pop III stars form either in isolation or in a binary, so we test the effects of this assumption by comparing our predicted SFRD to a model run with a fixed star-formation efficiency (SFE) of 0.001 in Figure~\ref{fig:SFRD_cosmo}. With a fixed SFE, more stars will form in more massive halos than less massive ones and, in general, more stars will form in all halos than in our fiducial model. As a result, we see that an increased SFE raises the global SFRD, though the effect is modest, with the deviation between the models reaching a factor of a few at the latest times (near the peak of Pop III star formation). 

For a more detailed discussion of the star formation process in individual minihalos we refer the reader to the analysis in M18, for which our model is analogous. Here, we instead now turn to an exploration of the variations in the global SFRD that result from variations in the individual star formation physics.

\begin{figure}
    \centering
    \includegraphics[width = \columnwidth]{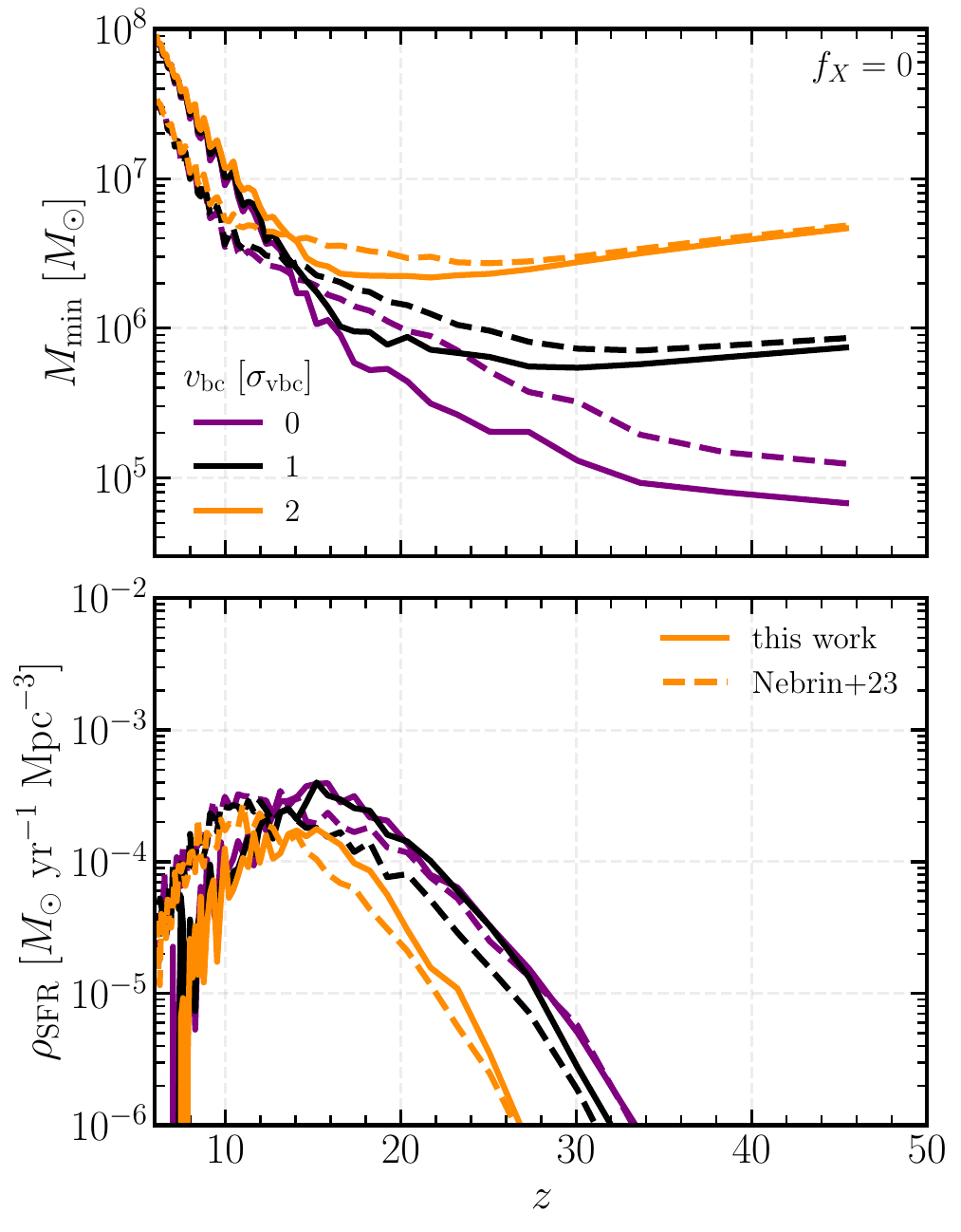}
    \caption{(top) The minimum mass for three different values of the stream velocity (different colors), comparing our calculation (solid curves) to that based on \citet{Nebrin23} (dashed curves; i.e., using their density threshold, taking $\zeta = 0.16$, and setting $\alpha_{\rm vbc} = 6$). (bottom) The associated Pop III SFRD for the same cases as above.}
    \label{fig:v_var}
\end{figure}

\begin{figure*}
    \centering
    \includegraphics[width=\textwidth]{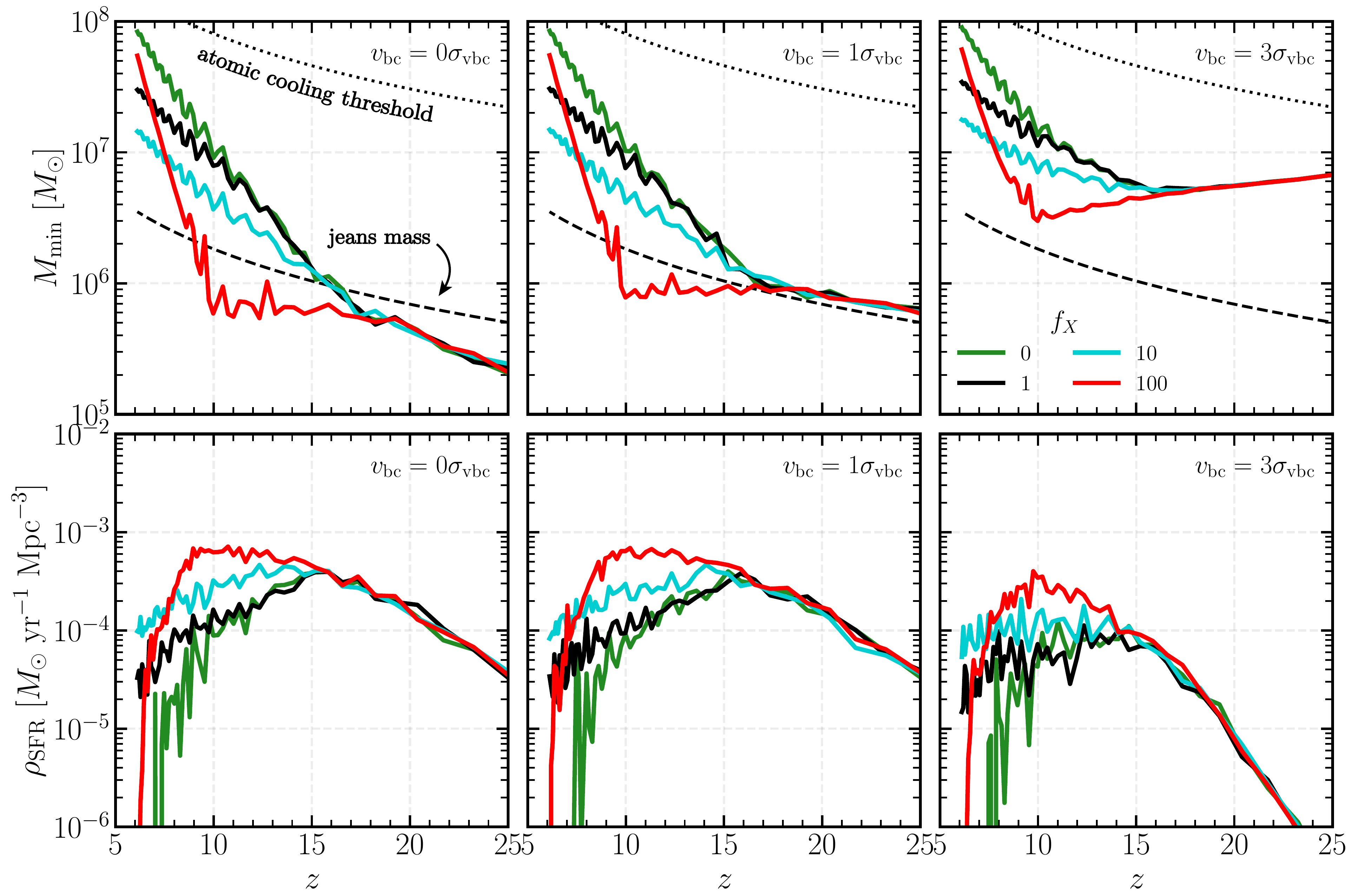}
    \caption{(top) The minimum mass as a function of redshift for $f_X=0, 1, 10, 100$ (green, black, blue, red, respectively) and $v_{\rm bc}=0,1,3\sigma_{\rm vbc}$ moving from left to right. The Jeans mass (effectively the minimum halo mass for the \textit{first} star to form; dashed black curve) and atomic cooling threshold ($T_{\rm vir} = 10^4\ {\rm K}$; dotted black curve) are shown for reference. (bottom) The associated Pop III SFRD for the same cases as the upper panels.}
    \label{fig:vpanels_fXvar}
\end{figure*}

\subsection{The effects of the minimum mass on Pop III star formation}\label{ssec:min_mass_variations}
The minimum mass model outlined in \S~\ref{sec:min_mass} allows us to naturally test the effects of varying the details of Pop III star formation physics. To this end, we examine the evolution of the minimum mass and Pop III SFRD for representative values of the stream velocity and X-ray background strength (which is parameterized by $f_X$; see equation~\ref{eq:SFR_emissivitty}) in Figures~\ref{fig:v_var} and~\ref{fig:vpanels_fXvar}. Note that in these figures, we have omitted the Pop II SFRD because the timing of the transition is effectively independent of the details of Pop III star formation (and our model assumes these halos are not subject to the same feedback mechanisms)---in other words, most halos transition to Pop II star formation once they cross the atomic cooling threshold, rather than as a result of metal enrichment. \footnote{For example, a $10^{13}M_\odot$ halo at $z=6$ will have crossed the atomic cooling threshold at $z\sim 43$, while for $10^{10}M_\odot$ and $10^8 M_\odot$ halos, that crossing point is $z\sim 23$ and $z\sim 10$, respectively.} We have also zoomed in on the times during which the effects of X-ray feedback will be dominant ($z\leq 25$).

First, we can isolate the effects of introducing and varying the stream velocity strength with no X-ray background (i.e., $f_X = 0$; Figure~\ref{fig:v_var}). We also demonstrate the effects of variations in our choice of halo gas density by overlaying the evolution of the minimum mass and SFRD using the fiducial parameters adopted in \citet{Nebrin23} (i.e., their density prescription, $\zeta = 0.16$, and $\alpha_{\rm vbc} = 6$; dashed curves). However, we defer discussion of those differences to \S~\ref{ssec:model_comparison} and focus on variations in $v_{\rm bc}$ and $f_X$ here. 

Broadly, the minimum mass is dominated by the cooling mass at early times and the joint LW-X-ray mass once star formation has built up for all values of $v_{\rm bc}$. As we increase the stream velocity, however, the filter and cooling masses grow in strength at early times and the minimum mass demonstrates a qualitatively different shape as a result (as it decays with the stream velocity $v_{\rm bc}\propto (1+z)$). Increasing the stream velocity from $0\to 1$ or $1\to2 \sigma_{\rm vbc}$ induces an increase in the minimum mass by a factor of $\sim 10-20$. This drastic increase in the minimum mass is clearly reflected in the Pop III SFRD for regions with a moderate or strong relative velocity ($v_{\rm bc} \gtrsim 2\sigma_{\rm vbc}$). Note that the effect of this in the SFRD is diminished somewhat in going from $0$ to $1\sigma_{\rm vbc}$ because of the additional requirement that a halo must build up enough gas to exceed the local Jeans mass (dashed black curve in Figure~\ref{fig:vpanels_fXvar}), which takes a significant time, thereby raising the effective minimum mass for the \textit{first} generation of star formation. Nevertheless, the first stars do not form until $z\sim 30-35$ in regions of the strongest relative velocity (compared to $z\sim 40-45$ in weak streaming regions)---a nearly 50 Myr delay in the onset of star formation. However, despite this delay, once minihalos begin to form Pop III stars, the SFRD quickly approaches the `equilibrium' level seen in the no/weak streaming case, so the late time behavior is indistinguishable from the weak streaming limit in most cases. In the case of strong relative velocity (yellow curves), the minimum mass is already growing steeply by the time the SFRD builds up to a high level, so it is not able to reach the same peak level as in regions with $v_{\rm bc}\lesssim 2\sigma_{\rm vbc}$.

In Figure~\ref{fig:vpanels_fXvar}, we isolate the effects of X-ray feedback at fixed $v_{\rm bc}$. First, across all three panels, it is evident that the inclusion of X-ray feedback significantly enhances Pop III star formation.
That is, moving from no X-ray feedback (green) to even a weak background (black) induces a factor of a few change in the minimum mass and flattens the redshift evolution somewhat. Indeed, the clearest difference between the runs with and without X-ray feedback are the duration of high levels of Pop III star formation---when X-ray suppression of the minimum mass is not present, Pop III star formation is quenched by $z\sim 7-8$.

Across all three streaming runs, both the positive and negative effects of X-ray feedback are apparent in the minimum mass. During the epoch where the minimum mass is set by LW feedback, positive feedback is manifested as a suppression to the minimum mass but the effect is relatively weak---the minimum mass is lowered by a factor of $\sim 5-10$ over two orders of magnitude increase in $f_X$. Despite this seemingly weak effect, the peak level of star formation achieved is boosted as we increase $f_X$. At the latest times, however, with $f_X = 100$ (red), the X-ray background has built up sufficiently for the IGM heating to translate into a steep increase in the filter mass, which can be seen in the smooth upturn of the minimum mass curves at $z\sim5-10$, where it switches back from tracing the LW or cooling curves to again following the filter mass. The timing of this transition is dependent on the strength of the X-ray background; with $f_X\gtrsim 50$ the IGM is efficiently heated (Figure~\ref{fig:xray_heat_ion}) after $z\sim 15$ and the minimum mass rises steeply in response. This negative feedback is reflected in the Pop III SFRD for the strongest X-rays---as heating overtakes the suppression due to enhanced positive X-ray feedback, $M_{\rm min}$ approaches the same late time value ($\sim 10^8 M_\odot$) as the no X-ray model and Pop III star formation is again shut off at the latest times, though the transition happens more sharply than the no X-ray case. 

Analyzing the effects of $v_{\rm bc}$ and $f_X$ together, it is apparent that the independent variations in the minimum mass and SFRD persist in the presence of both effects. That is, as we increase $v_{\rm bc}$, the peak SFRD is increasingly suppressed and as we increase $f_X$, it is boosted. In the case of the strongest X-rays and strong streaming (red curve in rightmost panels), X-ray heating is significant at nearly the same time that the LW mass takes over from the cooling threshold. The steeper growth of the minimum mass in this case results in the SFRD plummeting just as quickly as it grew. 

We note that, regardless of $f_X$ and the local stream velocity, all of these models yield a peak Pop~III SFRD $\sim 5\times 10^{-4} \ M_\odot$~yr$^{-1}$~Mpc$^{-3}$. These parameter choices primarily determine the duration of that peak era. However, the peak SFRD will be sensitive to other choices in the model; for example, we assume that each halo forms only one or two stars at a time. The SFRD will be directly proportional to the average mass of stars formed in each event.

\subsection{The effects of bursty Pop II star formation}\label{ssec:bursty_comparison}
\begin{figure}
    \centering
    \includegraphics[width=\columnwidth]{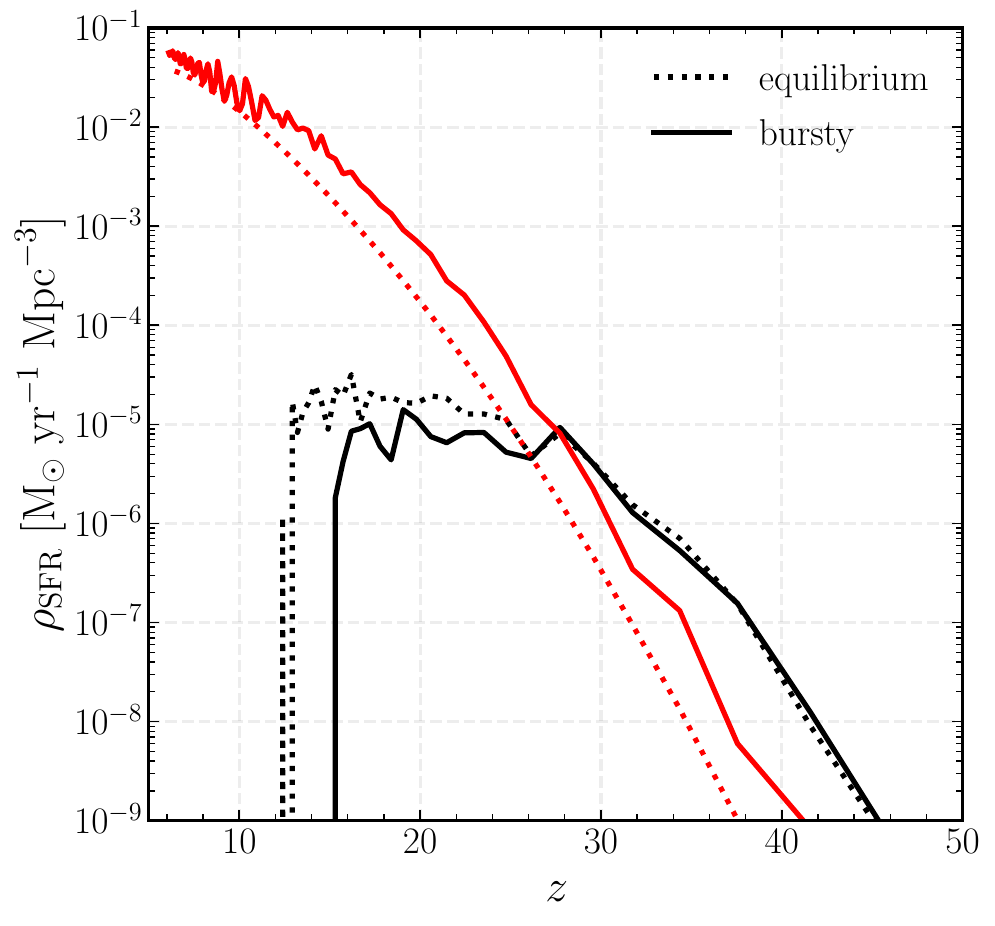}
    \caption{The Pop III (black) and Pop II (red) SFRD as a function of redshift comparing the bursty (solid) and equilibrium (dotted) models for the minimum mass prescription used in M18.}
    \label{fig:bursty_compare}
\end{figure}

In Figure~\ref{fig:bursty_compare} we explore the effects of bursty Pop II star formation on the global Pop III and II SFRD. For clarity in isolating the effects of bursty star formation, we present the SFRD calculated for the minimum mass prescription used in M18. As described in \S~\ref{sec:semianalytic_model}, cycles of bursty star formation should cause Pop II halos to overshoot the equilibrium SFR, as is reflected in the two red curves. The signature of these cycles is most clearly seen at low redshifts (when our 1 Myr timesteps are much more finely spaced in redshift space) but they affect the first Pop II halos as well. An increase in the Pop II SFRD leads to a quicker buildup of the LW background and thus an increase in the minimum mass. This in turn has the effect of suppressing Pop III star formation---the Pop III SFRD plateau achieved in the bursty case is smaller than that of the equilibrium case and star formation shuts off at a redshift of $\sim 16\ (12)$ in the bursty (equilibrium) case. In this sense, bursty Pop II star formation acts as another negative `feedback' mechanism for Pop III star formation. 

However, we include bursty star formation in our fiducial model and in the parameter variations shown in Figure~\ref{fig:vpanels_fXvar} and find that the negative effects are subdominant relative to the other effects setting the minimum mass. That is, the inclusion of X-ray ionization and updated models for self-shielding, for example, suppress the minimum mass to a level that bursty star formation in Pop II halos does not meaningfully affect the ability of a minihalo to form Pop III stars.     

\section{Comparison to other works}\label{sec:comparison}
Several other groups have considered the question of when and where Pop III stars can form, both from a numerical and analytic standpoint. Because this epoch of the universe's history has yet to be observed, the uncertainties are large and it is useful to consider a range of models that explore the entire parameter space. In this section we compare our results to those obtained by other groups, first focusing on the minimum mass calculation and next on the semi-analytic model. 

\subsection{Minimum mass model}\label{ssec:model_comparison}
The minimum halo mass for molecular hydrogen cooling has been studied in a number of contexts. One of the first attempts to analytically describe early star formation came in \citet{Tegmark1997}, wherein the collapse threshold argument was introduced to identify when $\molh$ cooling will become efficient. We base the broad structure of our photodissociation criterion---see \S~\ref{sssec:LW_bckd}---on this argument. This argument was shown to produce minimum masses that qualitatively agree with the results of simulations (e.g., \citealp{Machacek01}, \citealp{OsheaNorman08}, \citealp{Visbal14}) though in detail the results differ by a factor of a few. The effect of streaming on gas accretion has been studied many times as well; we have already compared our results to \citet{Naoz13} in \S~\ref{sssec:filter_w_stream}, demonstrating good quantitative agreement.
  
\begin{figure*}
    \centering
    \includegraphics[width=\textwidth]{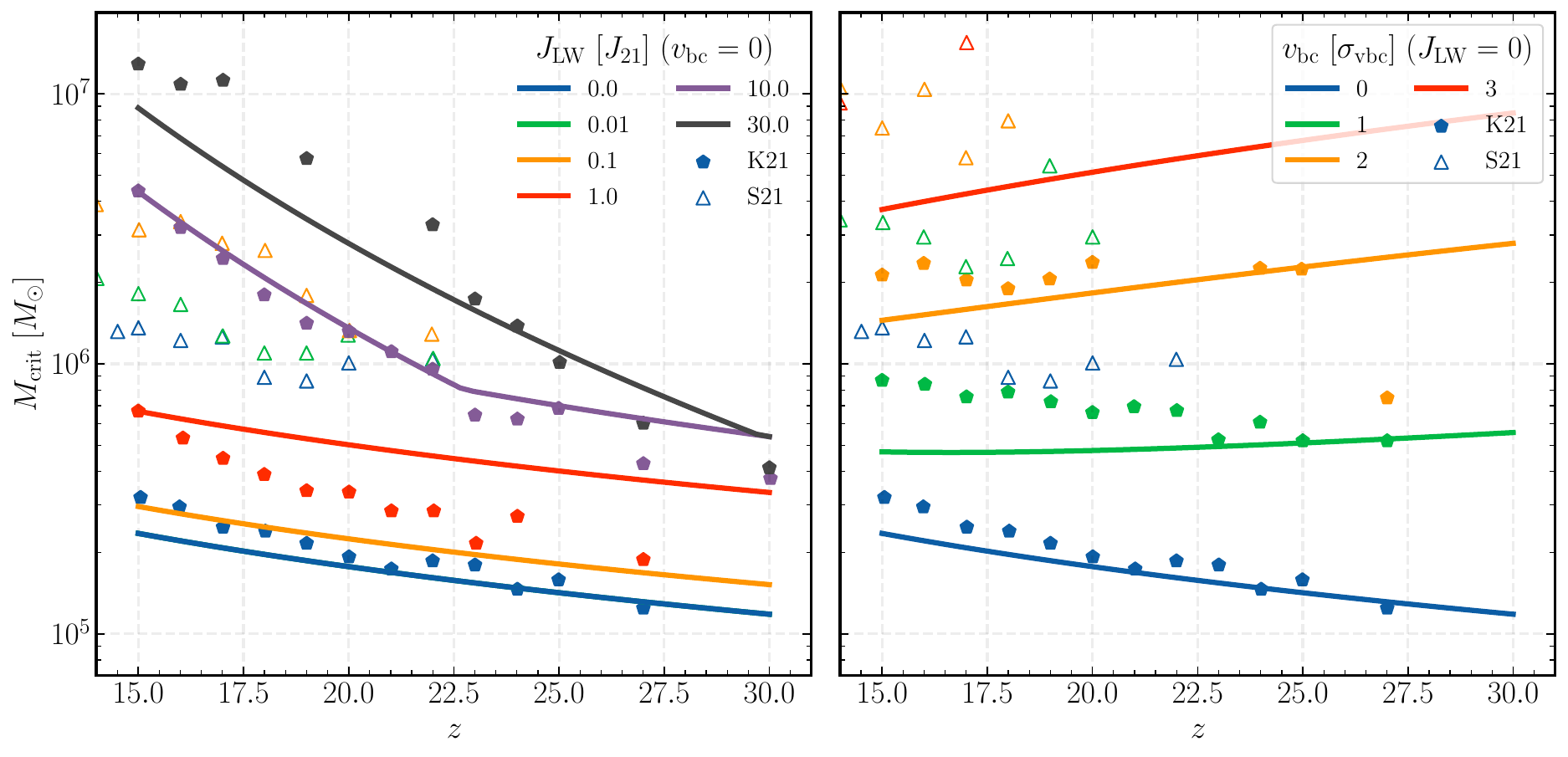}
    \caption{The redshift evolution of the minimum mass (solid colored curves) for various values of $J_{\rm LW}$ with no streaming (left) and for various values of $v_{\rm bc}$ with no LW background (right). These panels are the equivalent of Figures 2 and 3 in K21 and Figure 5 in S21. The simulated data from K21 (pentagons) and S21 (triangles) are overlaid in each case. Note that we have omitted the error bars on these data for readability. 
    For $J_{\rm LW} < 0.1 J_{21}$, the cooling mass is larger than the photodissociation threshold for all $z \geq 10$, so the blue and green curves lie atop each other in the left panel. }
    \label{fig:kul_scha_compare}
\end{figure*}

The first simulations to study the \emph{joint} effects of LW photodissociation and the stream velocity on the minimum mass were K21 and S21. In Figure~\ref{fig:kul_scha_compare}, we compare our calculations to the simulated results of K21 and S21, isolating the effects of photodissociation and streaming. These two simulations tested various values for the LW background and stream velocity with the goal of estimating the minimum DM halo mass for star formation in the presence of these two effects. However, their simulations probe different regions of the LW parameter space---the K21 group used values of $J_{\rm LW}/J_{21} = 0, 1, 10, 30$ and S21 used $J_{\rm LW}/J_{21} = 0, 10^{-2}, 10^{-1}$---and are discrepant in their limited overlap. Therefore, we remain agnostic about committing to either set of results and defer to our analytic calculation of the minimum mass.\footnote{During the era of LW dominance in the minimum mass ($z\lesssim 25$; see Figure~\ref{fig:fid_SFRD} and \S~\ref{sec:results}), the LW intensity is of order $J_{\rm LW}/J_{21}\sim \mathcal{O}(1-10)$, so the K21 results likely probe the more relevant parameter space for regulating Pop III star formation.} 
We find that we can more closely reproduce their results with a modification of $\zeta$ in our cooling threshold, perhaps in part because of differences in the definition of the \textit{critical} (K21) or \textit{average} (S21) mass thresholds. In particular, for the following comparison, we take $t_{\rm cool} < t_{\rm ff} = \zeta t_H\implies \zeta = 0.16$.

In Figure~\ref{fig:kul_scha_compare}, we show that our analytic calculations agree reasonably well with the K21 simulations and underpredict the S21 results in the case of no streaming and no LW background (where the mass scale should be set by the $v_{\rm bc}=0$ cooling threshold described in \S~\ref{sssec:cooling_threshold}). This discrepancy continues as we introduce a nonzero LW intensity and stream velocity.

First, we fix $v_{\rm bc} = 0$ and isolate the effects of varying $J_{\rm LW}$ (left panel). Comparing the simulated results, we see that those of S21 demonstrate a larger minimum mass for $J_{\rm LW}/J_{21}=0.1$ than those of K21 for $J_{\rm LW}/J_{21}=1$. We find that the redshift evolution of our results for $M_{\rm crit}$ agrees qualitatively well with those of K21, especially for large $J_{\rm LW}$. For $J_{\rm LW}/J_{21} = 1$, however, our calculations differ from that of K21 by a factor of a few. One key difference in the simulations of S21 and K21 is in their calculation of $f_{\rm shield}$. In particular, K21 use the same self-shielding criterion as in this work (WG11/19), which was introduced as a correction to that used in S21 (namely, \citealp{DB96}) for warm, dense gas. Previous calculations of $M_{\rm min}$ that ignored the effects of self-shielding (e.g., \citealp{Machacek01}, \citealp{OsheaNorman08}) predict minimum masses nearly an order of magnitude larger than those found in this work. We find that the minimum mass is fairly sensitive to the magnitude and redshift evolution of the density threshold which in turn feeds into the strength of the self-shielding effect and speculate that this is perhaps one source of the discrepancy between K21 and S21.

Next, we fix $J_{\rm LW} = 0$ and isolate the effects of streaming. We find that the calculation of the thermalization threshold (\S~\ref{sssec:thermalization_threshold}) with our choice of $\alpha_{\rm vbc} = 5$ produces excellent agreement with the simulated results of K21. For $v_{\rm bc} \neq 0$, our calculations underpredict those of S21 by a factor of a few, with the discrepancy increasing with increasing $v_{\rm bc}$.
However, given that K21 and S21 disagree by similar amounts, it is not clear how to resolve this difference.

Despite the small quantitative differences noted in the evolution of the minimum mass displayed in Figure~\ref{fig:kul_scha_compare}, we find that the resulting Pop III SFRD changes only modestly if we replace our minimum mass model with the fitting formula reported in K21. 

On the other hand, we find that the large variations in $M_{\rm crit}$ with $v_{\rm bc}$ reported by S21 do significantly change the Pop III SFRD, if we use their fitting formula. For example, with $v_{\rm bc} = 3\sigma_{\rm vbc}$ their procedure makes the minimum mass  so large that no Pop III stars form (cf. the star formation delay we see in Figure~\ref{fig:vpanels_fXvar}).

The impact of an X-ray background on the Pop III star formation process is similarly uncertain. \citet{Ricotti16} studied the joint effects of LW photodissociation and X-ray feedback analytically, using a cooling and Jeans mass criterion to set the minimum mass. They found that this criterion defined a global feedback loop, wherein Pop III stars can positively and negatively feed back into the star formation process, with the dominant effect depending on the strength of the X-ray background. Both of their conditions for defining the minimum mass are similar in spirit to ours but differ in detail, due to differences in the details of our cooling criterion, X-ray calculations, and our use of the filter mass rather than the Jeans mass. Qualitatively, however, we find a similar feedback loop wherein a critical value for the X-ray strength ($f_X\gtrsim 40$ for our model) defines the transition between positive and negative feedback dominance. The simulations of \citet{Park21} follow up on this analytic calculation to investigate the effects of incorporating local feedback processes in more detail and find broad qualitative agreement with the results of \citet{Ricotti16}.

\begin{figure}
    \centering
    \includegraphics[width=\columnwidth]{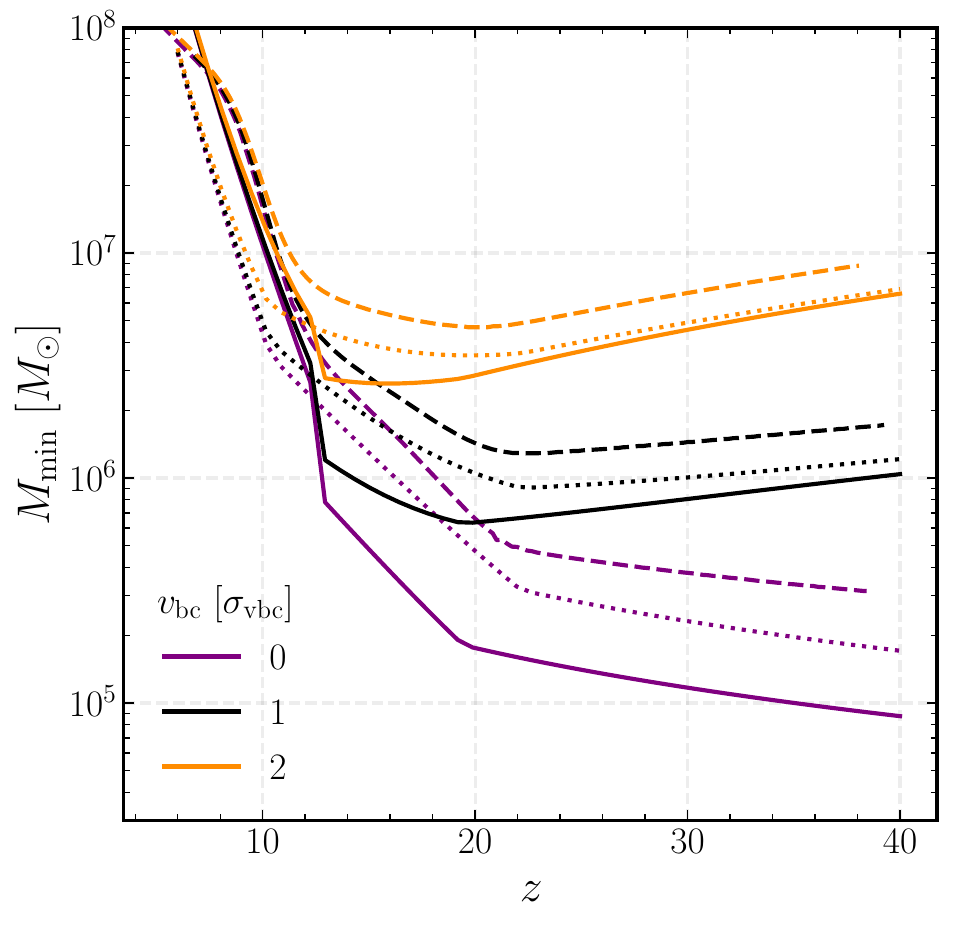}
    \caption{A comparison of the redshift evolution of the minimum mass for three representative values of the stream velocity ($0, 1, 2\sigma_{\rm vbc}$ --- purple, black, and orange, respectively) between our model and that of \citet{Nebrin23}. The three sets of curves compare our calculation (eq.~\ref{eq:overall_threshold}; solid curves), the \citet{Nebrin23} model (dashed curves), and our model with the \citet{Nebrin23} density criterion (dotted curves). This is analogous to their Figure 9. For ease of comparison, all take the Lyman-Werner radiation background presented in \citet{Incatasciato23}.}
    \label{fig:nebrin_compare}
\end{figure}

While this project was being completed, a related study developing a semi-analytic estimate for the minimum mass (akin to our \S~\ref{sec:min_mass}) was released \citep{Nebrin23}. In Figure~\ref{fig:nebrin_compare}, we compare our results to their minimum mass model, adopting the same Lyman-Werner background evolution \citep{Incatasciato23} and cooling threshold (i.e., $t_{\rm cool}< 6t_{\rm ff} = \zeta t_H\implies \zeta \simeq 0.16$) to simplify the comparison. While their model is fairly similar in spirit to that presented in this work, in detail, there are some key differences that result in distinct calculations of the minimum mass. For example, they consider the joint effects of $\molh$ cooling, LW feedback, and the stream velocity on the minimum mass, as we do. They include the effects of reionization feedback---which we do not include---though it is likely to be significant only at fairly late times ($z \lesssim 10$) and be patchy until reionization is complete (at $z \lesssim 6$). We do not include such inhomogeneous mechanisms here. They do not include the effects of an X-ray background, which we show to be important in both promoting and suppressing high redshift star formation between $z\sim 6-20$ (\S~\ref{sssec:xray-bkgd}). Therefore, we limit our comparison with their results to the overlap: the calculations of the cooling, photodissociation, and thermalization thresholds, for which there are some differences.

Based on the comparison in Figure~\ref{fig:nebrin_compare}, the similarities between our models are clear. That is, the redshift evolution is comparable and the minimum mass clearly transitions between the cooling mass and LW mass at roughly the same time in the two models. Despite the broad similarity, there is a difference of a factor of a few in the magnitude of the minimum mass at the highest redshifts (most notably in the case with no streaming). This difference can be attributed to their choice of central halo gas density (i.e., comparing our equation~\ref{eq:ngas} to their equation 7). As a result, our model has a critical $\molh$ cooling fraction that is two orders of magnitude smaller than their equation 16.\footnote{Evaluating our eq.~\ref{eq:fH2_crit} for $M = 10^6M_\odot$ and converting from $\zeta \to \eta$ at $z=9$, we find $f_{\molh, \rm crit}\simeq 2.08\times 10^{-5}$.} This also results in a somewhat different redshift evolution of the cooling mass (we find a different temperature scaling for the low-mass density threshold). This difference will also feed into the calculation of the dissociation threshold. Indeed, if we modify our density threshold to match theirs (dotted curves in Figure~\ref{fig:nebrin_compare}), much of the difference between our models at small streaming velocities is erased.

Our treatment of the contribution to the minimum mass due to the stream velocity is identical to that of \citet{Nebrin23} (i.e., we both employ the threshold identified in \citealp{Fialkov12}), but we also include the calculation of the filtering scale (accretion threshold) as a baseline for our model and choose $\alpha_{\rm vbc} = 5$ for our fiducial calculations (while they set $\alpha_{\rm vbc} = 6$). As such, we see similar evolution of the minimum mass with increasing $v_{\rm bc}$, though the magnitude of this increase differs for the same reasons as outlined above. Despite these differences, at late times our calculations converge to the same minimum mass value $M_{\rm min}\to 10^8 M_\odot$ as $z\to6$.

In Figure~\ref{fig:v_var}, we overlay the results of our semi-analytic calculation assuming the same fiducial parameters as \citet{Nebrin23} (dashed curves; i.e., with their density threshold, taking $\zeta = 0.16$, and setting $\alpha_{\rm vbc} = 6$) to test how these assumptions feed into the predictions made with our semi-analytic model. At high redshifts, the evolution of the minimum mass is qualitatively similar between the two cases, with the increased $\alpha_{\rm vbc}$ setting the discrepancy. As expected, this leads to a slightly earlier onset of star formation in our models than we would see with a stronger effect from the stream velocity. At late times, the differences between the models are a result of the different density parameterizations. In particular, our high-mass density threshold is normalized to a somewhat smaller value than that chosen in \citet{Nebrin23} (cf. we choose $n_{\rm H}\sim 12\ {\rm cm}^{-3}$ while they choose $n_{\rm H} \sim 32\ {\rm cm}^{-3}$ at $z=20$). With all else fixed, this means that, once we are in the high-mass regime, a halo needs to be more massive to approach the same critical dissociation threshold. Because the density evolves with redshift, this is equivalent to an earlier rise in the minimum mass to reach the same fixed density, which we see. Combining the two effects, the resulting SFRDs are effectively discrepant
by a constant offset of $\Delta z\sim 1$. While this would in turn produce different observable signatures, we emphasize that such discrepancies are significantly smaller than the uncertainties in other parameters outside of the minimum mass model, such as the selection of $f_X$ and the Pop III IMF.

\subsection{The Pop~III star formation history}\label{ssec:sam_comparison}

Next, we focus our comparison on two recent works studying semi-analytic modeling of the formation of the first stars: \citet{Visbal20} and \citet{Magg22}. We note that both of these works use the same LW feedback prescription used in M18, which does not include the effects of self-shielding and to which we compare our calculation in Figure~\ref{fig:fid_SFRD}. We demonstrate that inclusion of $\molh$ self-shielding can suppress the minimum mass by nearly an order of magnitude during this era and so expect that these results will overpredict the suppression of Pop~III star formation by the LW background (and hence underpredict the resulting SFRD). We also note that neither of these works includes X-ray feedback.

\citet{Visbal20} apply analytic models for star formation to DM halo merger trees from cosmological simulations to model the Pop III star formation process. This approach, combined with their grid-based calculation of the radiation backgrounds, allows them to account for local, inhomogeneous effects on the SFRD, such as clustering and mergers. Rather than following star formation in individual halos (as we do), they assume a fixed fraction (0.001) of the gas in a halo is converted into Pop III stars in a single star formation event, corresponding to roughly $100 M_\odot$ of stars in a $10^6 M_\odot$ minihalo. After a fixed delay time, these halos are assumed to have re-accreted metal-enriched gas and will transition to equilibrium Pop II star formation with a similar fixed star formation efficiency of 0.05. While qualitatively similar to our approach, their model differs in its details: we have a longer delay for re-accretion of enriched gas ($\sim 50$~Myr, c.f. their Fig.~3), and we treat the subsequent enriched star formation more carefully (without imposing a fixed star formation efficiency). But the most important difference is in our use of a self-shielding prescription. Ultimately, they find sustained Pop III star formation to $z=6$ (as we do), but find a Pop III SFRD nearly an order of magnitude smaller than our results, likely primarily because of differences in the minimum mass. In their model, the transition to Pop II star formation also occurs at earlier times than ours ($z_{\rm trans}\sim 25-30$ for our model). Nevertheless, the overall histories are qualitatively similar. 

\citet{Magg22} similarly develop their semi-analytic model using DM merger trees from cosmological N-body simulations with the goal of analyzing the effects of variations in the recovery time (time to transition to metal enriched star formation) on the 21-cm signal (see \S~\ref{sec:21cm}). In a procedure akin to that of \citet{Visbal20}, they assume a fixed star formation efficiency for Pop II stars and for their total mass of Pop III stars. However, they do not explicitly include the LW feedback or stream velocity in their calculation of the minimum mass, and they ignore X-ray feedback. Instead, they parameterize the minimum mass in terms of a critical temperature and calculate this threshold (which is based on the local circular velocity) using individual pixels in their 21-cm simulation box. They also employ a different IMF, choosing a power-law with $\phi\propto M^{0.5}$ (i.e., a very top-heavy IMF), compared to our Chabrier IMF, which has a power-law slope of $-2.35$ at large masses. Similarly to \citet{Visbal20}, they specify a single event of Pop III star formation, followed by a fixed recovery time before a 
transition to metal-enriched star formation (for which they consider several possibilities; our fiducial model corresponds most closely to their `intermediate' (30 Myr) transition case.

In this case, we find  qualitative agreement with their model, although their transition redshift is somewhat later than ours ($z_{\rm transition} \sim 20-25$ compared to $25-30$ in our fiducial model) and they find much higher peak levels of Pop~III star formation at a level 
about an order of magnitude larger than the
peak late-time value we see. These are likely a result of their much higher star formation efficiencies ($\approx 0.02$, which is a factor of 20 larger than the value chosen in \citealt{Visbal20},and corresponds to a $M_{\rm III} \sim 10^3 M_\odot$ starburst in the smallest halos). Because many more Pop III stars form in their model than ours, it takes longer for the Pop II star formation to catch up and they find a somewhat later transition redshift. Despite the differences, Pop~III star formation persists to late times in both of our calculations, though they find nearly an order of magnitude larger Pop III contribution to the SFRD than in our model. This illustrates the sensitivity of the SFRD to the assumed Pop~III star formation efficiency. Otherwise, the qualitative similarity of our histories suggest that it is the total Pop~III star formation efficiency, integrated over all star formation episodes, that sets the amplitude of the SFRD, whether or not a single event is assumed or repeated cycles (as found in the simulations of \citealt{Abe21}).

Finally, we compare our predictions to those of two high resolution cosmological simulations: \citet{Jaacks19} and \citet{Liu20}. 

\citet{Jaacks19} utilize detailed subgrid models in conjunction with cosmological simulations to predict the evolution of the global star formation rate density prior to the epoch of reionization. They self-consistently model the formation and evolution of both Pop III and Pop II stars and the transition between the two generations. Comparing the predictions of our fiducial model to theirs, we find good agreement in the expected SFRD. Though star formation turns on earlier in our semi-analytic calculations, we find that the overall SFRD approaches a similar level between our models ($\rho_{\rm SFR, peak}\sim {\rm few}\ \times 10^{-4} M_\odot\ {\rm yr^{-1}}\ {\rm Mpc^{-3}}$ at $z\sim 10$).

\citet{Liu20} combine cosmological simulations with semi-analytic models to study the global end stages of Pop III star formation. In particular, they consider the effects of growing radiation backgrounds, metal enrichment, and reionization to determine the magnitude of Pop III star formation following the reionization epoch. Given that the focus of this work is the evolution of the global star formation rate at high redshifts, it is most instructive to compare our outputs in the range of overlap: namely, between $z\sim 10-25$. Comparing our fiducial model to theirs, the qualitative structure is the same---Pop III star formation grows to a peak at $z\sim 10$ and steadily declines thereafter, largely in response to a growing LW background. However, the peak value achieved in their results ($\rho_{\rm SFR, peak}\sim 10^{-4} M_\odot\ {\rm yr^{-1}}\ {\rm Mpc^{-3}}$) is a factor of a few smaller than that which we find in our model, for the same reason as the differences noted with \citet{Visbal20} and \citet{Magg22}. That is, the use of a minimum star-forming mass scale for LW feedback calibrated to simulations that do not include the effects of self-shielding overestimate the effect of a growing LW background.

\section{Observational Implications}\label{sec:observations}
We extend our semi-analytic model to generate preliminary predictions of observable signatures of Pop III star formation in minihalos in the early universe. Unfortunately, direct observations of Pop III halos will be very challenging with both current and forthcoming telescopes. Indeed, M18 calculate that these halos will have absolute magnitudes between $M_{\rm AB}\approx-10$ and $-5$. Indirect signatures of Pop III stars---i.e., in their transients or in their effect on their surroundings---offer a far more promising avenue to constrain the physics of Pop III star formation with current technology. 

\begin{figure}
    \centering
    \includegraphics[width=\columnwidth]{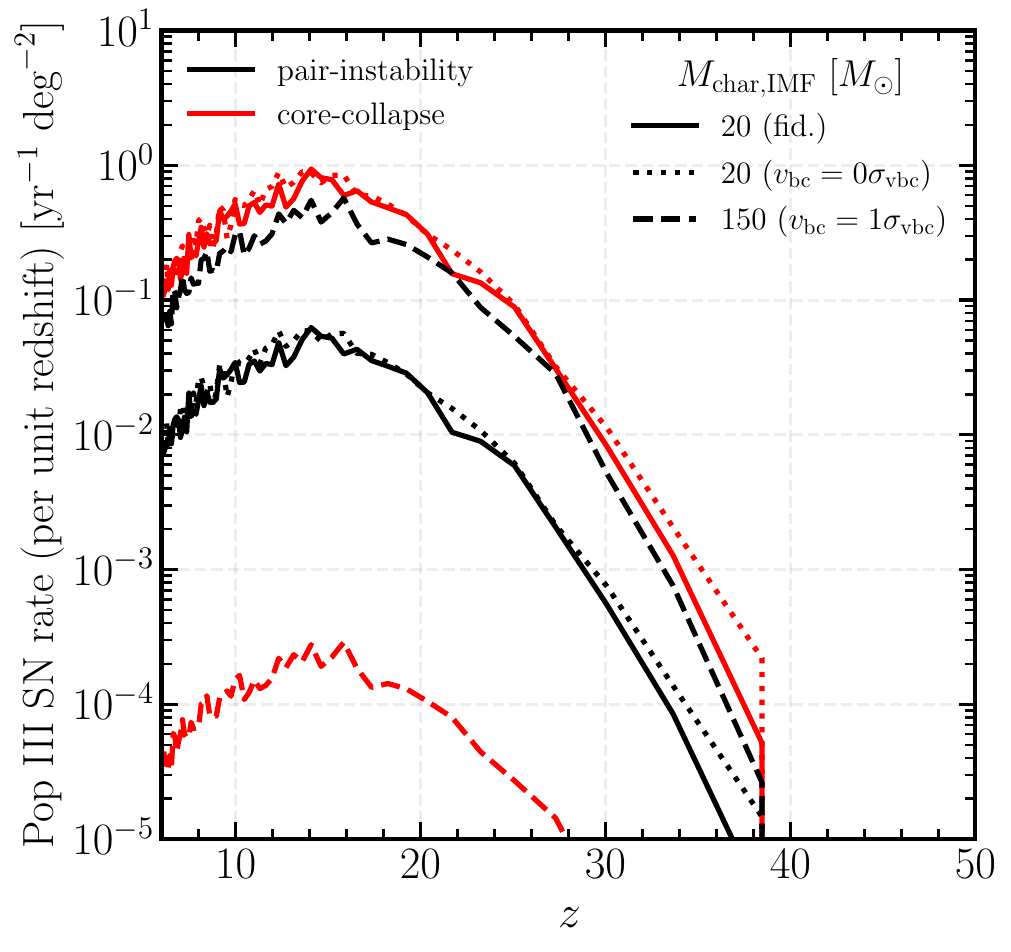}
    \caption{Redshift evolution of the Pop III SN rate for several models. We show results for our fiducial model (solid curves), the fiducial model in a region with no streaming (dotted curves), and a model with a much higher characteristic Pop~III mass (dashed curves). In each case, we show the PISN and core-collapse rates separately (black and red curves, respectively).}
    \label{fig:SN_rates}
\end{figure}

\subsection{Pop III supernovae}
Stars with masses between $140-260 M_\odot$ are expected to end their lives in pair-instability supernovae (PISNe; \citealp{Barkat67}, \citealp{Fryer01}, \citealp{HegerWoosley02}, \citealp{HegerWoosley10}). These superenergetic supernovae can produce nearly two orders of magnitude more energy than traditional core-collapse SNe and are so powerful that they completely tear the progenitor star apart, leaving no compact remnant behind. If Pop III stars have a top-heavy IMF---as we expect from numerical simulations---it is likely that some of them will form in this mass range and could produce PISNe. We can calculate the transient rate as:
\begin{equation}\label{eq:transient_rate}
    \frac{d^2N}{dt_{\rm obs}d\Omega_{\rm obs}}(z) = \frac{\eta_{\rm IMF}}{1+z}\frac{d^2V}{dzd\Omega_{\rm obs}}\rho_{\rm SFR}(z)
\end{equation}
where $d^2V/dzd\Omega$ is the differential comoving volume element, $\eta_{\rm IMF}$ is the number of progenitors per unit stellar mass (and is determined by the IMF), the factor of $1/(1+z)$ accounts for cosmological time dilation, and $\rho_{\rm SFR}(z)$ is the usual SFRD calculated from our semi-analytic model. From this, we can identify the rates associated with different types of SNe by integrating the IMF over our range of interest 
\begin{equation}\label{eq:eta_IMF}
    \eta_{\rm IMF} = \frac{\int_X m\phi(m) dm}{\int m \phi(m) dm}
\end{equation}
where the numerator is an integral over the range of progenitor masses for the transient of interest \citep{LazarBromm22}.

We summarize the results of this basic calculation in Figure~\ref{fig:SN_rates}. By comparing runs with two different characteristic masses in the IMF---moderately top-heavy (solid) and extremely top-heavy (dashed)---we demonstrate the sensitivity of the supernova rates to the details of the IMF. That is, if the Pop III IMF is truly very top-heavy, as is shown in the dashed curve, the PISN rate will be several orders of magnitude larger than the core-collapse SN rate. \citet{Jaacks19} also predict SN rates associated with Pop II and III stars in their calculations. Scaling our predictions to their reported observing area and calculating the cumulative SN rate integrated over redshift, we find roughly equivalent predictions for PI and CC SNe as we expect given the similarity of our predicted SFRDs.

Of course, these supernovae are only visible in very deep surveys. \citet{Kasen11} simulated the light curves of PISNe and found a wide range of expected luminosities (see also \citealt{Hummel12}). Their results suggest that surveys reaching limiting magnitudes of $m \ga 30$ will be required to identify these sources, which (depending on the PISN model) could reach events at $z \sim 10$--25. Our model demonstrates that, even in optimistic models, such surveys must span many square degrees to accumulate a significant number of events.
 
\begin{figure}
    \centering
    \includegraphics[width=\columnwidth]{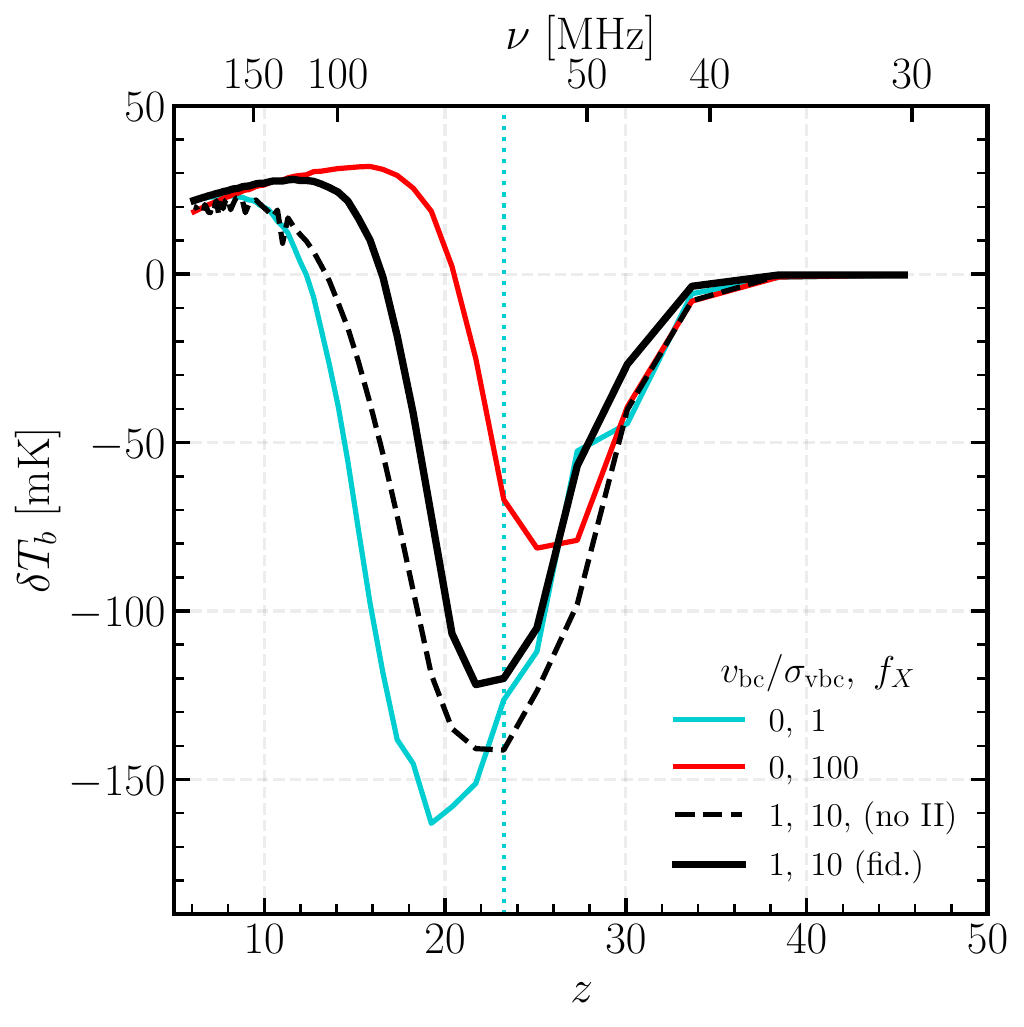}
    \caption{The differential 21-cm brightness temperature (relative to the CMB) for three representative Pop III star formation models---no stream velocity and weak X-rays (blue), no streaming and strong X-rays (red), and average streaming with moderate X-rays (the fiducial model; black)---and one with no Pop II star formation (dashed black). For each of these models, the transition from Pop III to Pop II dominance in the SFRD is indicated with a colored vertical dotted line.}
    \label{fig:21cm_Tb}
\end{figure}

\subsection{21-cm global signal}\label{sec:21cm}

Another potential indirect signature of Pop III stars is in their impact on their environment, the IGM. In particular, the sky-averaged 21-cm `spin-flip' signal of atomic hydrogen provides a tracer of the IGM that is sensitive to the electron fraction, temperature, and Ly$\alpha$ intensity. We have already demonstrated that high redshift star formation has the potential to very efficiently heat and ionize the IGM (Figure~\ref{fig:xray_heat_ion}). \citet{Mirocha18} and \citet{Mebane20} demonstrated that the inclusion of Pop III stars in calculations in the 21-cm signal can introduce a characteristic asymmetry in the absorption trough due to early star formation. 

Here we extend our semi-analytic model to include calculation of the Ly$\alpha$ background following \citet{Holzbauer12}:
\begin{equation}
    J_\alpha(z) = \sum_{n=2}^{n_{\rm max}}\int_z^{z_{\rm max}(n)}dz' f_{\rm recycle}(n)\frac{(1+z)^2}{4\pi}\frac{c}{H(z')}\epsilon(\nu_n', z')
\end{equation}
where $z_{\rm max}$ is the maximum redshift that a Ly$n$ photon can experience before it redshifts into the next transition, and $\nu_n'$ is defined as above (equation~\ref{eq:ion_rate}). We sum over Ly$n$ levels because a fraction $f_{\rm recycle}(n)$ of Ly$n$ photons will cascade down to become Ly$\alpha$ photons via a series of radiative transitions (\citealp{Hirata06}, \citealp{PritchardFurlanetto06}). In practice, we truncate this sum at $n=23$ to exclude the levels for which the photon horizon lies within the size of a ${\rm H\ II}$ region of a typical galaxy at this time. We calculate the emissivity with a procedure akin to our calculation of the LW emissivity:
\begin{equation}
    \epsilon(z) = \int_{M_{\rm min}}^\infty n(M) \frac{\Omega_b}{\Omega_m}\frac{\dot{M}_*}{m_p}\bigg(\frac{N_{{\rm Ly}n}E_{{\rm Ly}n}}{\Delta \nu_{{\rm Ly}n}}\bigg)dM
\end{equation}
where $M_{\rm min}$ is the minimum mass for star formation, $\dot{M}_*$ is the star formation rate of a halo of mass $M$, $N_{{\rm Ly}n}$ is the number of Ly$n$ photons produced per baryon in stars \citep{BarkanaLoeb05}, $E_{{\rm Ly}n}$ is the average energy of a Ly$n$ photon, and $\Delta \nu_{{\rm Ly}n}$ is the frequency spacing between the $n$ and $(n+1)$st Lyman line. 

The spin temperature of the 21-cm line is set by a competition between scattering of CMB photons, collisions, and scattering of Ly$\alpha$ photons (\citealp{Wouthuysen52}, \citealp{Field58}, \citealp{FurlanettoOhBriggs06}). Following \citet{Furlanetto06} and \citet{Mebane20} (and the references therein), we calculate the coupling coefficients $x$ to compute the spin temperature
\begin{equation}
    1-\frac{T_\gamma}{T_S} = \frac{x_c+x_\alpha}{1+x_c+x_\alpha}\bigg(1-\frac{T_\gamma}{T_K}\bigg)
\end{equation}
and from this the 21-cm brightness temperature (relative to the backlight of the CMB)
\begin{equation}
    \delta T_b = 27 x_{\rm HI}\bigg(\frac{\Omega_b h^2}{0.023}\bigg)\bigg(\frac{0.15}{\Omega_m h^2}\frac{1+z}{10}\bigg)^{1/2} \bigg(\frac{T_S-T_\gamma}{T_S}\bigg)\ {\rm mK}
\end{equation}
where $x_{\rm HI}$ is the neutral fraction.

With this, we compute the differential 21-cm brightness temperature for three representative models (Figure~\ref{fig:21cm_Tb}). Note that we have \textit{not} included the ionizing UV background that will drive the process of cosmic reionization, so these calculations are meant to be diagnostics of the \textit{early universe} signatures of Pop III physics on the 21-cm global signal. We have also ignored the potential effects of a cosmic radio background from accreting black holes (see e.g., \citealp{Mebane20}, \citealp{Ventura23}).

Indeed, the shape and timing of the signal are clearly sensitive to the global processes that govern the SFR. When star formation is strongly coupled to the X-ray luminosity (red curve; strong X-rays), the IGM is efficiently heated and the depth of the absorption trough is suppressed. As a result, the signal turns over to emission more quickly, yielding a very sensitive probe of the X-ray background strength. In kind, there is a clear signature of the stream velocity in the width and timing of the absorption trough. That is, with the stream velocity, star formation is delayed and the absorption trough in turn is delayed as well. At late times, however, the star formation rate quickly climbs to match the non-streaming levels and the signals are indistinguishable. 

For $z\gtrsim 20-25$, the contributions of Pop III and II stars to the SFRD (and thus global signal) are comparable. Indeed, comparing our fiducial model to one with no Pop II contribution to the radiation backgrounds (dashed black), the depth and shape of the absorption trough are different at late times (following the transition to Pop II star formation). At early times, Pop III stars tightly couple the spin temperature to the gas kinetic temperature driven by the Wouthuysen-Field effect. At later times, the spin temperature remains tightly coupled to the temperature of the IGM, but the enhanced SFRD (in response to the growth of Pop II halos) drives more rapid heating of the IGM and thus the absorption signal is somewhat suppressed. 

Comparing these calculations to the results of \citet{Mebane20}, we find higher-redshift features than seen before. Namely, the updated calculation of the LW mass introduces an earlier onset of the absorption trough than seen in \citet{Mebane20}. We find that variations in the X-ray background strength can suppress the depth of the absorption trough, as they do.

\citet{Ventura23} also predict the 21-cm global signal using a semi-analytic model for high-redshift star formation. Though there are differences in our modeling of the star formation process, the overall evolution of the 21-cm signal that they calculate is in good agreement with that which we see in our fiducial model in Figure \ref{fig:21cm_Tb}.

We can also compare our 21-cm global signal predictions to those made in \citet{Magg22}. We note that they neglect to include the effects of X-rays (which will dominate the heating of the IGM) in their calculation of the 21-cm signal. As expected, we find a similar absorption depth and qualitative global signal to their `intermediate' transition model. However, we find that the nadir of the absorption trough is achieved at higher redshifts in our model ($z_{\rm abs}\sim 20-25$ compared to 10-15 in their model) and find a characteristic high redshift contribution to the absorption from Pop III stars that is not reflected in their results. This discrepancy is puzzling because we see comparable (or larger) contributions to the Pop III SFRD between their model and ours (see \S~\ref{ssec:sam_comparison}). We speculate that this is perhaps a result of the lack of X-ray heating included in their model. For example, if we compare the weak and strong X-ray curves (blue and red in Figure~\ref{fig:21cm_Tb}, respectively), we see that the inclusion of a strong X-ray background moves the nadir of the absorption trough earlier and suppresses its depth. In \citet{Magg22}, the only source of heating is from Ly$\alpha$ photons, whereas we include both contributions, so heating in their model proceeds more slowly.

Though this calculation makes several simplifying assumptions, this estimate demonstrates that variations in the Pop III star formation physics are reflected in the 21-cm global signal, especially at the highest redshifts. Indeed, in all of the displayed cases, we find the characteristic asymmetry in the absorption trough produced by early IGM heating from high-redshift Pop III star formation identified by \citet{Mirocha18}. Forthcoming low-frequency radio telescopes, such as the lunar FarView array, plan to study the early universe through the highly redshifted 21-cm line. Measurements of the global signal at frequencies below 70~MHz will offer a strong and promising probe of the details of Pop III star formation physics.

\section{Conclusions}\label{sec:conclusions}
We have presented a simple analytic model for the minimum DM halo mass for Pop III star formation that incorporates the combined effects of a relic relative DM-baryon `stream' velocity from the early universe and feedback from the buildup of UV/LW and X-ray backgrounds. Such a  criterion is the crux of any semi-analytic model for high-redshift star formation and allows us to self-consistently model the formation of the first stars and the transition to subsequent generations of metal-enriched star formation. We incorporate this analytic calculation, a Chabrier-like IMF, and a criterion for `bursty' Pop II star formation as updates to the semi-analytic model of M18. 

From this model, we identify three key epochs of Pop III star formation. At the earliest times ($z\gtrsim 30$), the stream velocity is the dominant environmental factor in setting the Pop III SFRD---so the SFRD is most sensitive to choice of $v_{\rm bc}$ in this era. However, the stream velocity decays as $v_{\rm bc} \propto (1+z)$, so for intermediate times ($30 \gtrsim z\gtrsim 10$), the LW background sets the Pop III SFRD, coincident with the transition to Pop II star formation dominating the SFRD. Finally, at the latest times ($z\lesssim 10$), the persistence of Pop III star formation is dictated by the strength of the X-ray background. While bursty star formation can suppress Pop III star formation through the LW background, we find that this is a secondary effect to those of the aforementioned three processes.

These distinct epochs in the minimum mass are translated into variations in the Pop III SFRD. In particular, the onset of early star formation is directly responsive to the strength of the DM-baryon relative velocity and the late-time magnitude of the SFRD (i.e., the peak of the SFR) is sensitive to the physics of the LW and X-ray backgrounds. Therefore, observations that even indirectly probe the Pop III SFRD will be a powerful tool for understanding the physics governing star formation in that era. 

While Pop III halos will likely not be directly observable with current (or near-future) technology, indirect signatures of these stars are a promising avenue to probe this epoch. We find that superluminous Pop III SNe are theoretically observable with forthcoming deep JWST and RST surveys and can shed light into the IMF for these first stars. Similarly, global 21-cm experiments that are able to reach the lowest frequencies will offer us a window into Cosmic Dawn and will potentially be able to constrain the physics of Pop III star formation. Recent work has shown that Pop III star formation is not limited to Pop III halos alone; accretion of pristine gas from the IGM can result in Pop III star formation in Pop II halos, though these stars will likely be very rare and short-lived \citep{Maio11_feedback, Venditti23}.

Although our model includes most of the key physical processes regulating Pop~III star formation, we have ignored processes for which small-scale inhomogeneities  are essential (such as metal enrichment and photoheating from UV photons during reionization). In the future, incorporating these processes will be essential for modeling spatial fluctuations in the Pop~III population.

\section*{Data Availability}

No new data were obtained as part of this work. Results used to generate the figures are available from the authors upon reasonable request.

\section*{Acknowledgements}
The authors thank Matt McQuinn, Rick Mebane, Smadar Naoz, Claire Williams, William Lake, Massimo Ricotti, and Roy Zhao for useful discussions. This work was supported by the National Science Foundation through award AST-1812458. In addition, this work was directly supported by the NASA Solar System Exploration Research Virtual Institute cooperative agreement number 80ARC017M0006. 

\textit{Software}: \textsc{numpy} \citep{numpy}, \textsc{astropy} \citep{Astropy}, \textsc{matplotlib} \citep{Matplotlib}, \textsc{scipy} \citep{Scipy}

\bibliographystyle{mnras}
\bibliography{biblio} 

\bsp
\label{lastpage}
\end{document}